\documentstyle[aps,prb,multicol,epsf,graphicx]{revtex}

\newcommand \nn{\nonumber}        
\newcommand \be {\begin{equation}}
\newcommand \ee {\end{equation}}
\newcommand \bea {\begin{eqnarray}}
\newcommand \eea {\end{eqnarray}}

\newcommand \bi {\bibitem}
\newcommand \s {\sigma}
\newcommand \lan {\langle}
\newcommand \ran {\rangle}

\begin{document}

\title{Order-parameter fluctuations (OPF) in spin glasses: Monte Carlo
simulations and exact results for small sizes}

\author{Marco Picco}
\address{{\it LPTHE}\\
       \it  Universit\'e Pierre et Marie Curie, PARIS VI\\
       \it Universit\'e Denis Diderot, PARIS VII\\
        Boite 126, Tour 16, 1$^{\it er}$ \'etage, 4 place Jussieu\\
        F-75252 Paris CEDEX 05, FRANCE\\
E-Mail: picco@lpthe.jussieu.fr}
\author{Felix Ritort and Marta Sales}
\address{Departament de F\'{\i}sica Fonamental, Facultat de F\'{\i}sica,
         Universitat de Barcelona\\
         Diagonal 647, 08028 Barcelona (Spain)\\
E-Mail: ritort@ffn.ub.es,msales@ffn.ub.es}

\date{\today}

\maketitle

\begin{abstract}
The use of parameters measuring order-parameter fluctuations (OPF) has been
encouraged by the recent results reported in \cite{RS} which show that
two of these parameters, $G$ and $G_c$, take universal values in the
$\lim_{T\to 0}$. In this paper we present a detailed study of
parameters measuring OPF for two mean-field
models with and without time-reversal symmetry which exhibit different
patterns of replica symmetry breaking below the transition: the
Sherrington-Kirkpatrick model with and without a field and the Ising
p-spin glass (p=3). We give numerical results and analyze the
consequences which replica equivalence imposes on these models in the
infinite volume. We give evidence for the transition in each system
and discuss the character of finite-size effects. Furthermore, a
comparative study between this new family of parameters and the usual
Binder cumulant analysis shows what kind of new information can be
extracted from the finite $T$ behavior of these quantities. The two main 
outcomes of this work are: 1) Parameters measuring
OPF give better estimates than the Binder
cumulant for $T_c$ and even for very small systems they give evidence
for the transition. 2) For systems with no time-reversal symmetry,
parameters defined in terms of connected quantities are the proper
ones to look at.
\end{abstract} 



\section{INTRODUCTION}

One of the main features of the low-temperature phase of spin-glasses is
the lack of self-averageness of the order parameter. Parameters measuring
order-parameter fluctuations (OPF), $A$, $A_c$, $G$, and $G_c$, where
introduced to locate phase transitions in systems which did not contain
time-reversal symmetry (TRS) in the Hamiltonian~\cite{MNPPRZ}. Very
recently the use of these parameters has been enhanced in~\cite{RS}
(henceforth referred to as RS) where it has been shown that $G$ and $G_c$
take, in the low-temperature limit, the universal values $1/3$ and $13/31$
respectively, provided the ground state is unique and that there is no gap
at zero field. This situation is generically met in spin-glasses with no
gap at zero coupling, thus bringing more interest to the applicability of
these parameters in the study of phase transitions.

In ordered systems, a good parameter to locate phase transitions is the
Binder cumulant which is the kurtosis of the order-parameter probability
distribution. The {\em uniqueness} of the ground state in that case is
enough to guarantee that the Binder cumulant takes the universal value 1 at
zero temperature for any finite volume. On the contrary, in order to find
universal values for $G_c$ and $G$, one has to add a condition, namely the
absence of gap in the local field distribution.

Concerning disordered systems, it is well known that the use of the Binder
cumulant may present serious problems, specially in models without
TRS in their Hamiltonian. Then, the Binder
cumulant may take negative values showing no evidence for a crossing of the
different curves for different sizes. Recently, Hukushima and Kawamura
\cite{POTTS} have shown that this is specially true in models with one-step
replica symmetry breaking where the Binder cumulant may take a negative
value just below $T_c$. This explains the non-positiveness of the Binder
cumulant in Potts glasses and $p$-spin models. The same being also true for 
$p$ even when there is time-reversal symmetry \cite{CCP}.

However, at finite $T$, parameters which measure OPF always remain finite
and positive. In systems with non-vanishing OPF, the behavior of these
parameters in the thermodynamic limit can be inferred from the restrictions
that the replica equivalence property imposes on the $P(q)$, $q$ being the
overlap between two different replicas of the same system. Even though the
behavior of $G$ is trivial, the behavior of $G_c$ is not, and depends on
which kind of replica symmetry breaking pattern (RSB) characterizes the
frozen phase. This knowledge can help us to distinguish among different
systems and can also provide us with useful information about the spin-glass
(SG) phase. 

One of the main outcomes of RS is that already very small sizes may yield
valuable information about the infinite volume spin-glass transition. This
is confirmed by previous numerical studies of Ising spin glasses and p-spin
models \cite{MNPPRZ,PPR,BCF}, Migdal-Kadanoff spin glasses \cite{BBDM},
Heisenberg spin glasses \cite{HK}, Potts glasses \cite{POTTS} and some
protein folding models \cite{PPRICCI}. In view of the main results
obtained in RS we will concentrate our attention on models where the
new parameters $G$ and $G_c$, as well as $A$ and $A_c$ may yield valuable
information not attainable using the Binder cumulant analysis.  In
particular, we are interested in models without time-reversal symmetry such
as spin glasses in magnetic field and p-spin models.

Our purpose is to present a detailed numerical analysis of parameters which
measure OPF by studying very small samples. Numerical calculations have
been performed using exact computations of the partition function for small
samples and Monte Carlo simulations.  Monte Carlo calculations have been
done in the spirit of the study on the spherical Sherrington-Kirkpatrick
(SK) model shown in RS where very small sizes were simulated. One of the main
outcomes of RS is that calculations of $G$ and $G_c$ at low-temperatures
need a very large number of samples. This is due to the fact that at low
temperatures only those samples having local fields much smaller than $T$
give a significant contribution to $G$ and $G_c$, so that the uncertainty
in determining the value of $G$ and $G_c$ becomes very large. In other
words, the low-temperature behavior of $G$ and $G_c$ provides direct
information about the quality of the data. Having values for $G$ greater
than $1/3$ shows either a lack of thermalization or poor sample
statistics, a situation which the Binder cumulant does not reflect, since
it is much less sensible to these fluctuations.

The paper is organized as follows. The next section presents the
definitions used in this paper. Section III analyzes the consequences the
replica equivalence imposes on the values of these parameters at finite
$T$. Sections IV and V analyze two mean-field models exhibiting different
kinds of RSB: the SK model with and without a magnetic field and the Ising
p-spin model with p=3. Finally, in section VI, we present the
conclusions. Two appendices are devoted to several technical points.

\section{A short reminder}
Order-parameter fluctuations are an intrinsic characteristic of disordered
systems. Although extensive quantities are self-averaging the order
parameter may not. This is the typical situation found in systems where due
to RSB there is a multiplicity of states and ergodicity is broken. Also in
the case where replica-symmetry is not broken OPF are very sensitive to the
spectrum of fluctuations around the stable solution. Spin glasses and
disordered systems in general are characterized by the presence of large
scale domain excitations without a typical length scale, and as a result
correlations have a very slow decay, typically as a power law, and hence,
with no characteristic length scale~\cite{}. The equilibrium phase is then
marginally stable and OPF decay slower than $1/V$ where $V$ is the
volume. The possible scenarios concerning the behavior of OPF have been
discussed in RS. Here we content ourselves with defining the parameters of
interest in the analysis of this paper.

Four parameters which measure OPF were studied in RS: $G, G_c, A,
A_c$, where the subindex $c$ stands for connected disorder quantities (to
distinguish from disconnected quantities). $B$, the usual Binder 
cumulant and $B_c$, its connected counterpart were used in \cite{PR} 
to study the four dimensional spin glass with binary couplings. 
Calling $q={1\over V} \sum_{i} S_i^a S_i^b $ the spin overlap of two
systems of spins $S_i^a$ and $S_i^b$, with $V$ the number of spins, we
define these parameters as follows,  

\be 
B=\frac{1}{2}\left(3-\frac{\overline{\langle q^4\rangle}}{\overline{\langle
q^2\rangle}^2}\right) \; , 
\label{eq2bb}
\ee

\be
A=\frac{\overline{\langle q^2\rangle^2}-\overline{\langle q^2\rangle}^2}
{\overline{\langle q^2\rangle}^2} \; ,
\label{eq2}
\ee

\be
G=\frac{\overline{\langle q^2\rangle^2}-\overline{\langle q^2\rangle}^2} 
{\overline{\langle q^4\rangle}-\overline{\langle q^2\rangle}^2}=\frac{1}{2}
\frac{A}{1-B}\label{eq1} \; ,
\ee

\be
B_c=\frac{1}{2}\left(3-\frac{\overline{\langle  (q-\langle
q\rangle)^4\rangle}}{\overline{\langle  (q-\langle
q\rangle)^2\rangle}^2}\right) \; ,
\label{eq4bb}
\ee

\be
A_c=\frac{\overline{\langle (q-\langle
q\rangle)^2\rangle^2}-\overline{\langle (q-\langle q\rangle)^2\rangle}^2} 
{\overline{\langle (q-\langle q\rangle)^2\rangle}^2} \; .
\label{eq4}
\ee

\be
G_c=\frac{\overline{\langle (q-\langle
q\rangle)^2\rangle^2}-\overline{\langle (q-\langle q\rangle)^2\rangle}^2} 
{\overline{\langle (q-\langle q\rangle)^4\rangle}-\overline{\langle
(q-\langle q\rangle)^2\rangle}^2}=\frac{1}{2}\frac{A_c}{1-B_c} \; . 
\label{eq3}
\ee
All of these parameters vanish at high temperature. It is well known that
in the limit $T\to 0$ the Binder ratio 
takes the universal value $1$, provided the ground state is unique. In
\cite{RS} it was shown that under the hypothesis of a unique ground state
and no gap in the local field distribution these parameters satisfy the
following properties,
\be
\lim_{T\to 0}G(V,T)=\frac{1}{3}
\label{eq5a}
\ee
\be
\lim_{T\to 0}G_c(V,T)=\frac{13}{31}\label{eq5b}
\ee
\be
\lim_{T\to 0}A(V,T)=0 \; .
\label{eq6}
\ee
The last identity for $A$ follows trivially from (\ref{eq2}) due to the
uniqueness of the ground state. In Appendix A we make a brief summary of
the main outcomes of section VI in RS to show how these universal values
are obtained.  The low-temperature results for $B_c$ and $A_c$ are also
derived in that appendix. The results are: 
\be
\lim_{T\to 0}A_c(V,T)\rightarrow\; \infty ~~~~~, ~~~~\lim_{T\to
0}B_c(V,T)\rightarrow \;-\infty \; . 
\label{eq6bb}
\ee These last two relations follow from the fact that even though
numerator and denominator for both quantities vanish at zero
temperature, the denominator vanishes as a higher power of $T$,
and thus the ratio diverges.

We have to note that $A$ and $A_c$ have a non trivial behavior so they
must be analyzed in detail in order not to extract wrong conclusions
at low temperatures as shown by B. Drossel et al. in a very recent
study of these parameters in the Migdal-Kadanoff
approximation~\cite{DBM}. In particular, one has to be careful when
dealing with transitions at $T=0$, as the limits $T\to 0$ and
$V\to\infty$ do not commute, so that results in (\ref{eq6bb}) could
not hold.

All these quantities are dimensionless, thus we expect that a
finite-size scaling analysis can be performed \cite{PPR}: 

\be X=f_X\left(L(T-T_c)^{1/\nu_X}\right) ~~~~ X~ =~ G,\,A,\,B \; .
\label{eq2Ds}
\ee Quantities which can be obtained from derivatives of the free
energy, {\it i.e.} thermodynamic quantities, are expected to have the
same critical exponents, hence, for the Binder cumulant we have that
$\nu_B=\nu$ (the usual correlation length exponent). However, we do
not know {\em a priori} how to express $A,G$ in terms of thermodynamic
quantities, so we cannot assure that scaling exponents for $G$ will be
the same as the ones for $B$. The equivalent of (\ref{eq2}) defined in
terms of the magnetisation instead of $q$, has been extensively
studied by numerical simulations for diluted models \cite{MADRID} and
for the Ashkin-Teller random model \cite{WD}. In both cases the
exponents associated with the parameters $A$ and $B$ ($G$ is only
function of $A,B$, see eq. (\ref{eq1})), within numerical precision, have been
shown to be the same. Here we reach similar conclusions for the SK
model.

Despite that critical behavior for $A,B,G$ is governed by the same
exponents ({\it i.e.} by the critical behavior of the same correlation length)
the parameters which measure OPF appear to be good indicators for a
phase transition. In the spin-glass phase $G$ and $G_c$ attain a finite
value and vanish in the paramagnetic phase. On the contrary, $A$ may
vanish in both the spin-glass phase and paramagnetic phase but stays
finite at $T_c$ (a result predicted for diluted systems
\cite{AH}). The interest of studying $A$ is that it does not vanish when
OPF are finite, a situation characteristic of replica symmetry breaking
transitions.  Such information can not be deduced only from $G$, $G_c$,
$B_c$ and $A_c$ which may be finite even when OPF vanish\cite{BBDM,RS}.

Let us stress that only in cases where time-reversal symmetry is absent
both class of parameters (connected or disconnected) can be studied. In
case of models with time-reversal symmetry in the Hamiltonian only
disconnected averages ($G,~A, ~B$) are feasible. A calculation of the
disconnected parameters $G_c,A_c$ and $B_c$ would require to artificially break
time-reversal symmetry which may impose additional difficulties to the
simulation. This explains why in \cite{RS} only the parameters $G$ and $A$
were studied for the Ising spin glass chain and the spherical SK model.

\section{Replica equivalence}

Systems exhibiting RSB at the transition are well-known to
have a non trivial $P(q)$ below the transition temperature. The information
about this probability distribution is contained in the replica matrix
$Q_{ab}$, where sub-indices stand for different replicas. The replica
equivalence property (RE)~\cite{PARISI1,BMY,MPSV} states that every
row/column of this matrix is a permutation of any other row. Thus, any
quantity such as $\sum_b Q_{ab}$ does not depend on $a$ or,
equivalently, the free energy must contain terms proportional to $n$
(which stands for the number of replicas) in order to ensure that there
will be no divergences when performing the limit $n \to 0$
\cite{PARISI2,AC}.

This property imposes some constraints on the $P(q)\equiv
\overline{P_J(q)}$. One can derive general relations for the probability
distributions of several overlaps and, as is pointed out in \cite{PARISI1},
any probability function of $m$ overlaps $P(q_{12},q_{34},q_{56}
,... ,q_{2(m-1) 2m})$ can be expressed in terms of the probability
distribution of one overlap $P(q_{12})$ plus the cyclic distributions, {\it
i.e.}  $P^{12,23,31}$,$P^{12,23,34,41}$,.. $P^{12,23,34,...,m1}$.  For the
case of two overlaps, fluctuations satisfy the following relation:

\be
P(q_{12},q_{34})=\frac{1}{3}P(q_{12})\delta(q_{12}
-q_{34})+\frac{2}{3}P(q_{12})P(q_{34}) \; ,  
\label {GUERRAb}
\ee
which in the thermodynamic limit yields the well-known Guerra relations
\cite{GUERRA1}: 
\be
\overline{\langle q_{12}^2\rangle^2}=\frac{1}{3}
\overline{\langle q_{12}^4\rangle}+\frac{2}{3}\overline{\langle
q_{12}^2\rangle}^2 \; ,
\label{GUERRA}
\ee
which, as stressed in RS yield for systems with non-vanishing OPF
(\ref{GUERRA}) a trivial behavior of  $G$, {\it i.e.}
\be
G=\frac{1}{3}\Theta(T-T_c) \; .
\label{gg}
\ee In RS it was pointed out that this result could hold even in
spin-glass systems with a marginally stable replica symmetric
phase. Notwithstanding, there is no such general trivial behavior for
the other disconnected quantities, $A$ and $B$, neither for $G_c$, $A_c$
or $B_c$. To compute $G_c$ one has to deal with the probability of
functions depending on three and four overlaps, so that in general one
has a much more complicated object which depends on the following terms:
\begin{eqnarray}
\frac{1}{n(n-1)}\sum_{a,b} Q_{ab}^k &=& \;\int P(q)q^k dq \; ,\\
\frac{1}{n(n-1)(n-2)}\sum_{a,b,c} Q_{ab}^kQ_{bc}Q_{ca}&=&\;
\int P(q_{12},q_{23},q_{31}) 
q_{12}^kq_{23}q_{31}dq_{12}dq_{23}dq_{31}\\
\frac{1}{n(n-1)(n-2)(n-3)}\sum_{a,b,c,d} 
Q_{ab} Q_{bc} Q_{cd} Q_{da}&=&\int P(q_{12},q_{23}q_{34},q_{41}) 
q_{12}^kq_{23}q_{34}q_{41}dq_{12}dq_{23}dq_{34}dq_{41}\nn \\
&& \equiv \frac{TrQ^4}{n(n-1)(n-2)(n-3)} \; .
\end{eqnarray}
Therefore, we expect that at finite $T$ the behavior of $G_c$ will
depend on the kind of RSB that takes place at the transition, as it depends
strongly on the structure of $Q_{ab}$.  When computing the connected
quantities, $A_c$ and $B_c$ one encounters the same situation so that we
have to go to particular patterns of RSB in order to obtain a simpler
expression for $G_c$, $A_c$ and $B_c$, as well as for $A$ and $B$.\\ 
In particular, for the case of one-step RSB, where $Q_{ab}$ can take two
different values $q_0$ and $q_1$, and the $n$ replicas are divided in $m$
blocks, we obtain a very simple expression for all the connected quantities
(calculations which lead to these results are shown in Appendix B): 
\be
G_c=\frac{39 - 113 m + 98 m^2}{93 - 221 m + 266 m^2} \; ,
\label{onestepgc}
\ee     
\be
A_c=\frac{39 - 113m + 98m^2}{140m(1-m)} \; ,
\label{onestepac}
\ee     
\be
B_c=\frac{3(31 - 167 m + 182 m^2)}{280 m (1-m)} \; .
\label{onestepbc}
\ee     
These results establish that connected quantities depend only on the RSB
parameter $m$, which in general is a function of $T$. However, $A$ and $B$
depend also on the elements of the replica matrix, namely $q_0$ and $q_1$,
as shown in the expressions (\ref{onestepa}) and (\ref{onestepb})
in Appendix B.

The simple functional dependence on $m$ of $A_c$, $G_c$ and $B_c$ lets us
establish several universal features of these parameters. In first place,
we note that below $T_c$, $G_c$ always remains finite and positive and that
at $T=0$, since $m=0$ in any system, it takes the universal value
$13/31$~\cite{RS}. In the same way, at $m=1$, $G_c$ takes another universal value: $G_c=0.174$. Another interesting feature  of  $G_c$ is the   
existence of a local minimum at $m=0.650$ whose value $G_c=0.113$ is
universal for all the systems exhibiting one-step RSB. As we will show in a
forthcoming section, this is important from the point of view of numerical
results for small sizes as the asymptotic position of this minimum can give
us information about the behavior in the thermodynamic limit and thus
about the dependence of $m$ on $T$.

In second place, we note that both $A_c$ and $B_c$ diverge at $m=0$, and
therefore at zero temperature, the same being also true at $m=1$. The
difference relies on the fact that while $A_c$ remains {\it always}
positive and has a minimum at $m=0.56$ whose universal value is
$A_c=0.187$, $B_c$ diverges to $-\infty$ and displays a maximum at a
positive value $B_c= 0.316~(\;m\;=\;0.451)$.  This is an interesting
result, on the one hand because $G_c$, $A_c$ and $B_c$ provide us with
valuable information about the characteristics of the SG phase, and on
the other, because the exclusive dependence of the connected quantities
on $m$ establishes a direct connection with models for structural
glasses, where $m$, the RSB parameter, is well-known to control the
violation of the fluctuation-dissipation theorem (FDT).

Unfortunately, for the two-step RSB scheme (and further RSB), we do
not have such a simple expression and numerical computations for each
system have to be performed.  In the two following sections we analyze
two mean-field models: the SK model in a field and the $p$-spin
$(p=3)$ Ising model. Neither of the two models contains TRS in the
Hamiltonian and exhibit RSB of different kinds: a full step and
one-step RSB respectively. Therefore, these systems are characterized
by having non-vanishing OPF and can provide a direct check of the
results obtained in this section.

\section{The Sherrington-Kirkpatrick model with and without a magnetic
field}

This section studies the Sherrington-Kirkpatrick model with and without
a field. Preliminary results for the model in a field were shown in
\cite{MNPPRZ}. For the model without a field results have been recently
presented by Hukushima and Kawamura \cite{POTTS}.  Here we present a more
systematic and detailed study for small sizes of connected and disconnected
quantities.  The SK model in a field is defined by the disordered
mean-field Hamiltonian,
\be
{\cal H}_{SK}=-\sum_{i<j}\,J_{ij}\,\s_i\,\s_j-h\sum_i\,\s_i \; .
\label{HSK}
\ee
Before discussing the results in a field we discuss the case of zero
field. An important difference between the study of OPF parameters and
Binder cumulants (for very precise recent SK simulation results see
\cite{CMPR,MZ}) is that very high-precision statistics is needed to
compute OPF parameters. Consider for instance $G$ and $G_c$. These are
ratios of two quantities which may be very small if OPF vanish yielding
a large error for $G$ and $G_c$. In models where OPF vanish simulating
large sizes may then require a prohibitive computational effort.  Not
only long simulations are needed to thermalize the samples but
high-precision statistics is needed to determine with reasonable
precision the ratio of quantities which vanish in the infinite-volume
limit. This second limitation is not present in models where OPF are
finite such as the SK model at finite temperature.

\subsection{Zero-field simulations}

\begin{figure}[tbp]
\begin{center}
\includegraphics*[width=12cm,height=8cm]{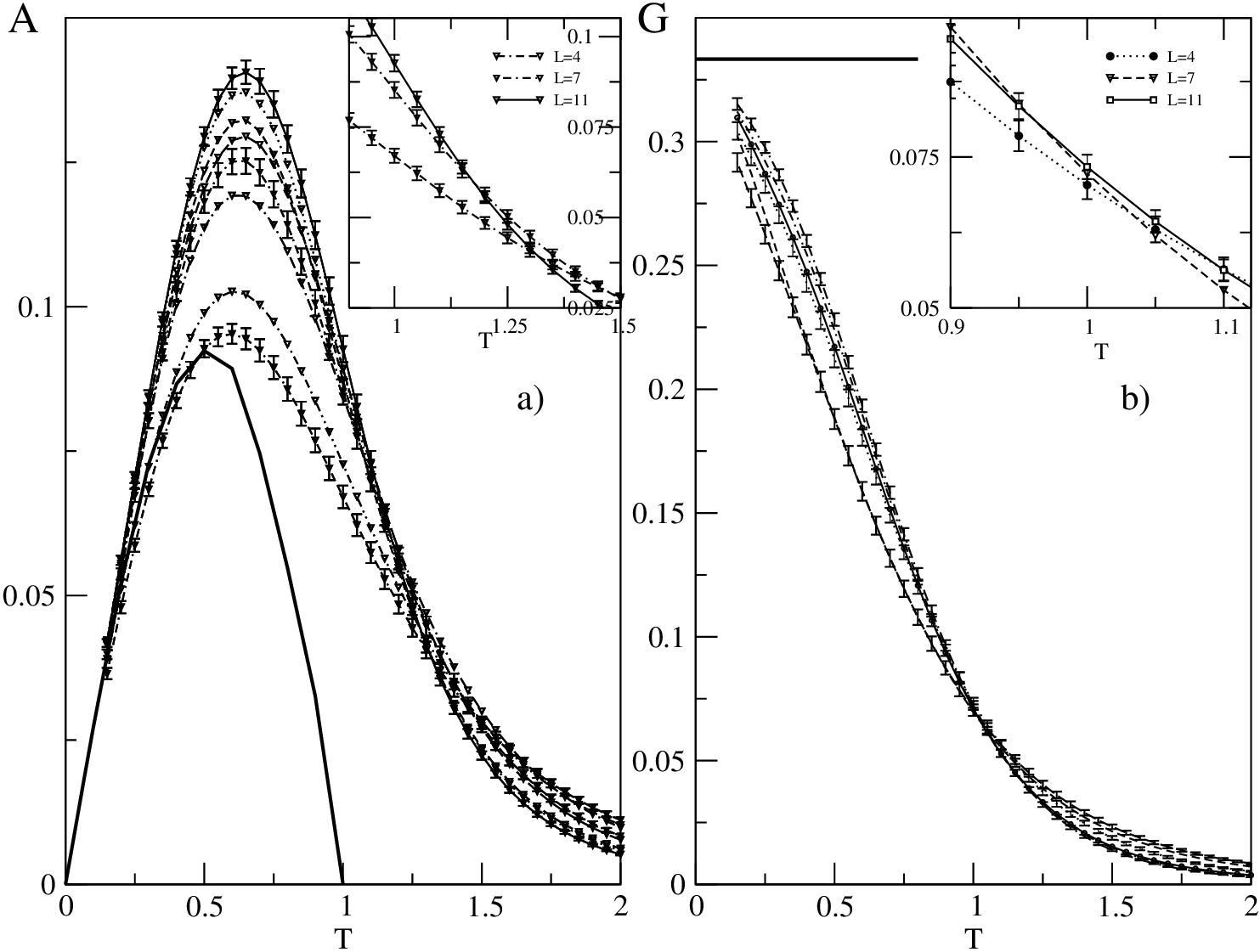}

\vskip 0.05in
\caption{Parameters $A$ (a) and $G$ (b) for the SK model with no field by
exact computations of the partition function averaged over $10000$ samples
for $N=4, 5, 6, 7, 8, 9, 10,11$ from bottom to top at low $T$. Error bars
are shown 
for sizes $N=4,7,11$. In both insets we show in detail the crossing region
for sizes $N=4,7,11$. The solid line in (a) corresponds to numerical
results of the one-step approximation~(\ref{onestepa}), obtained from
solving numerically the saddle-point equations of the SK model for this
particular case~\cite{BOOKS}.
\label{SK_h0_GA}}
\end{center}
\end{figure}

\begin{figure}[tbp]
\begin{center}
\includegraphics*[width=12cm,height=8cm]{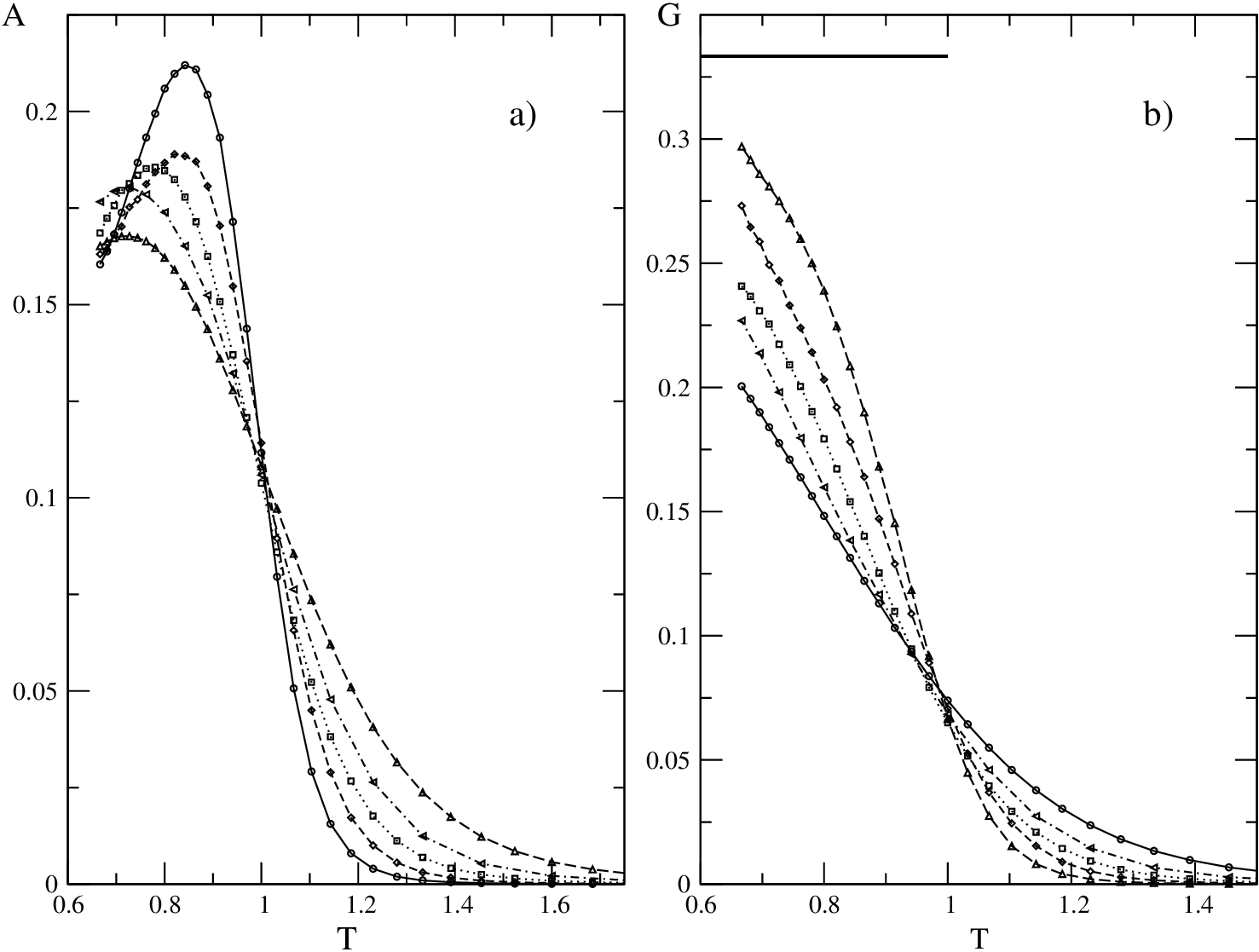}
\vskip 0.05in
\caption{Parameters $A$ (a) and $G$ (b) for the SK model with no field for
the case of binary couplings for sizes $N=32, 64, 128, 512$ from bottom to
top in the high $T$ region.  
\label{SK_h0_monte_GA}}
\end{center}
\end{figure}

\begin{figure}[tbp]
\begin{center}
\includegraphics*[width=12cm,height=8cm]{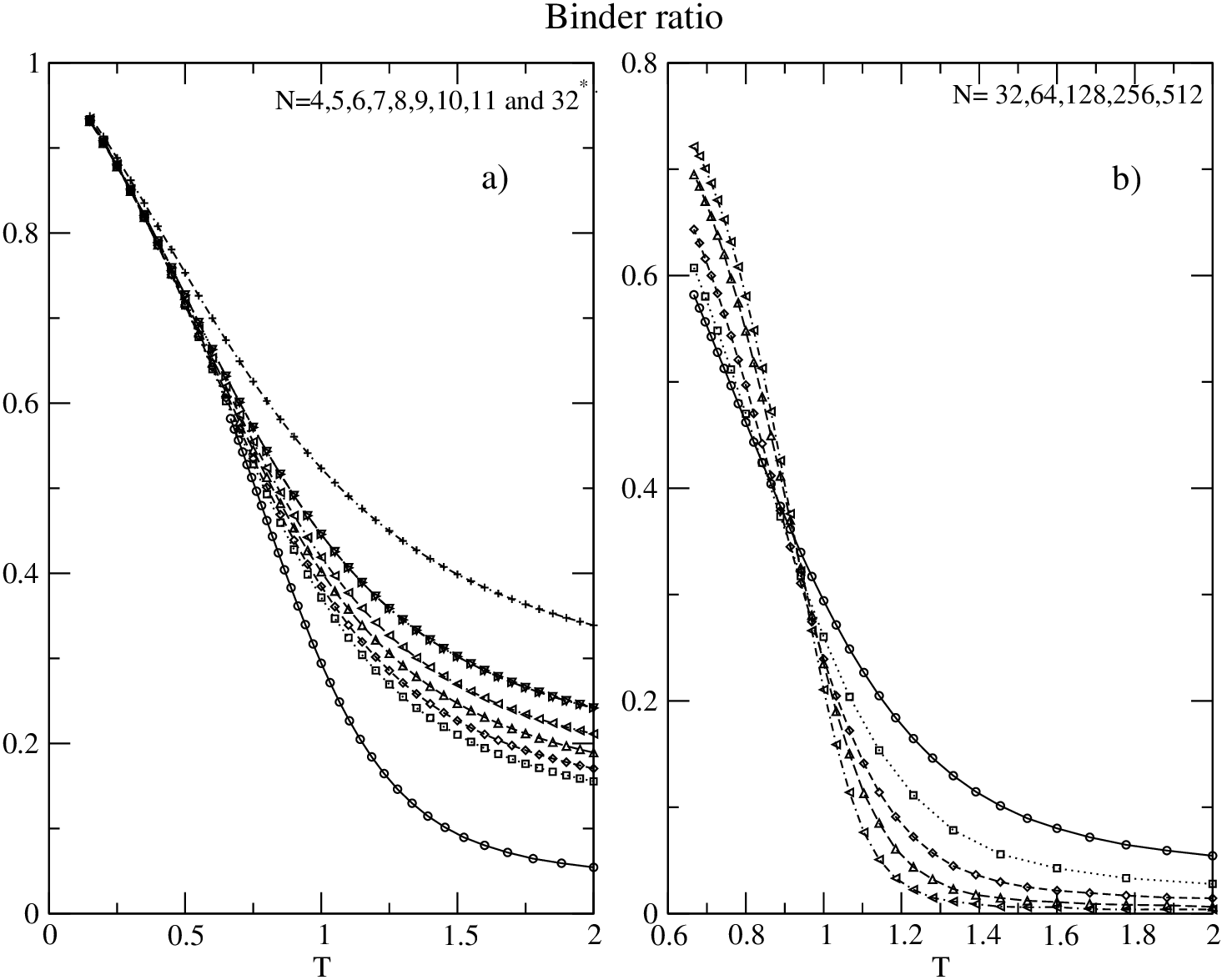}
\vskip 0.05in
\caption{Binder ratio,  $B$, for the SK model with $h=0$: (a) for small
systems ($N=4,5,6,7,8,9,10,11$ from top to bottom) and (b) for $N=32,
64,128,512$ from top to bottom at high $T$. 
\label{SK_h0_b}}
\end{center}
\end{figure}

\begin{figure}[tbp]
\begin{center}
\includegraphics*[width=12cm,height=8cm]{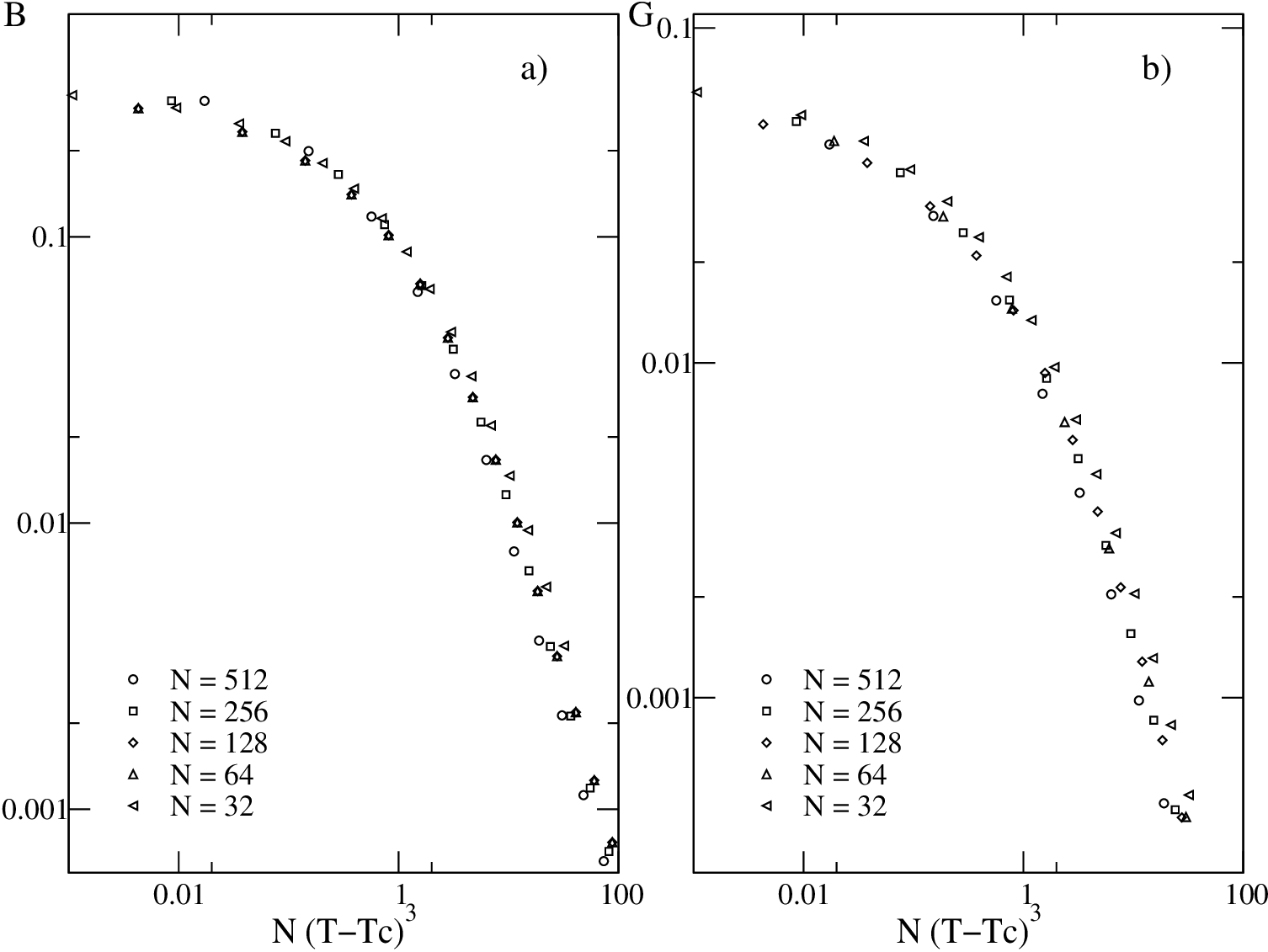}
\vskip 0.05in
\caption{Scaling functions of $B$ (a) and $G$ (b) for the SK model with
$h=0$ for sizes $N=32,64,128,512$. 
\label{SK_h0_sc}}
\end{center}
\end{figure}

Results for $G$ and $A$ are shown in figures \ref{SK_h0_GA} and
\ref{SK_h0_monte_GA}. In figure \ref{SK_h0_GA}, we display results for
very small sizes ($N=4\,-\,11$) obtained from exact computations of the
partition function (such analysis was done in \cite{YK} for
the SK model) averaging over $10000$ samples of Gaussian quenched
couplings with variance $1/N$.  In figure~\ref{SK_h0_monte_GA} we show
results for larger sizes ($N=32,64,128,256,512$) using the parallel
tempering technique \cite{HN,MPRRZ} with binary couplings. The number of
samples ranges from $1000$ for the smallest size, $N=\,32$, to $250$ for
the biggest one, $N=\,512$.  Figure \ref{SK_h0_b} shows the Binder ratio
$B$ for both groups of sizes. In figure \ref{SK_h0_sc} we show the
results of the scaling analysis performed to the results for $B$ (a) and
$G$ (b) for the larger sizes.

Monte Carlo simulations for larger sizes corroborate the results for $G$
and $A$ reported in \cite{POTTS}. It is important to note that, as this
system contains TRS, the Binder cumulant for different sizes is expected to
exhibit a crossing of the curves for different sizes as reported in
\cite{POTTS,BY}. However we must note that, while even for very small
samples we do observe a crossing for $G$ and $A$ at $T$ close to $T_c=1$
(figure \ref{SK_h0_GA}), for the same sizes, curves for $B$ do not display
such a crossing (figure \ref{SK_h0_b} (a)). In figure \ref{SK_h0_b} we
have plotted the Binder ratio for $N=32$ together with the smaller sizes to
stress that we have to increase the size of the system ({\it i.e.} reach
$N\approx\,100$) to observe the crossing and not only a merging of all the
curves in the low $T$ region. Results for larger systems display a crossing
of the curves of all the parameters as expected. Our observations
corroborate previous results~\cite{BCF} which argued that finite-size
corrections to $T_c$ for parameters measuring OPF were of opposite sign
than for the Binder parameter. For $G$ and $A$
(figures~\ref{SK_h0_GA} and ~\ref{SK_h0_monte_GA}) the crossing point at
$T^*$ starts well above $T_c$ and approaches $T_c$ as we increase the size
of the system. For the biggest sizes $T^*$ is very close to
$T_c$. On the contrary, curves for $B$ cross at $T^*<T_c$, and therefore,
since thermalization is more difficult at low $T$, the crossing is harder to
observe. Moreover, the crossing for the biggest sizes takes place at a
temperature slightly below $T_c$, thus making it evident that finite-size
corrections to $T_c$ are much stronger in the Binder ratio than in $A$ and
$G$.

The finite-size scaling analysis shown in figure~\ref{SK_h0_sc} shows a
remarkable result. In the particular case of the SK model, the argument
of the scaling function (\ref{eq2Ds}) for the Binder ratio can be shown
\cite{SPR} to be $N\,(T-T_c)^{3}$. The collapse of the data, which are
shown only for $T-T_c>1$, is very good, not only for $B$ but also for
$G$ (the same being true for the parameter $A$, although not reported
here). This suggests that exponents for $G$ are the same as for $B$,
justifying a search for a general field-theoretical argument in the line
of that proposed for diluted systems \cite{AH}.

As far as the details of the calculations are concerned, we have to point
out that for small systems Gaussian couplings assure the uniqueness of the
ground state, so that as reported in RS we obtain the universal values for
$G$ and $G_c$. For binary couplings, however, this results should not hold
for small systems, but as long as we increase the size of the system
({\it i.e.} we approach the thermodynamic limit) we should not expect a
dependence on the bond distribution because in mean-field entropy vanishes
at $T=0$ also in models with discrete couplings, and therefore, universal
results should also hold. In order to be sure that results can be compared,
we have checked that already for $N=33,32$ results for both coupling
distributions nearly coincide.

\subsection{Finite-field simulations}
In a field one must be a bit careful. According to \cite{RS} two conditions
are necessary to ensure equations (\ref{eq5a}) and (\ref{eq5b}): uniqueness
of the ground state and no gap in the local fields distribution. For the SK
model with Gaussian couplings these conditions are immediately
satisfied. But the first condition may be violated in the SK in a
field. This happens in samples which have a ground state with zero
magnetization. In this case the ground state is twofold degenerated because
both the zero magnetization configuration and its fully reversed one have
the same energy. In computing OPF for finite volume these rare samples have
a finite probability violating the requirement of {\em uniqueness} of the
ground state. Consequently eq. (\ref{eq5b}) for $G_c$ is not satisfied. Note
that, on the contrary, eq. (\ref{eq5a}) is still satisfied because it is
invariant against time-reversal symmetry. A zero magnetization ground
state can only occur for samples with an even number of spins, so we expect
strong differences for cases where $N$ is even or odd for small systems.

We have performed simulations with two applied fields $h=0.3$ and
$h=0.6$.  The transition into the frozen phase of the SK model in a
field is determined by the de Almeida-Thouless line \cite{BOOKS} which
yields $T^*_{h=0.3}=0.65$ and $T^*_{h=0.6}=0.48$.  In
figure~\ref{SK_h03_G}, we show results at field $h=0.3$ for an odd
number of spins for sizes $N=3,5,7,9,11$ by exact calculation of the
partition function and $N=13,17,23,33,64$ by Monte Carlo
simulations. The largest size ($N=64$) has been simulated with binary
couplings and parallel tempering, while the smallest ones have been
simulated by simple Metropolis algorithm with Gaussian distributed
$J_{ij}$. As we have said at the beginning of this section, when systems
are small and the ground state is not unique, eq. (\ref{eq5b}) is
violated as can be observed in the inset of the plot for $G_c$
(figure~\ref{SK_h03_G} (c)) where $G_c$ becomes greater than $13/31$,
and goes to $1$ at $T=0$. This does not happen when systems are small
and there is an odd number of spins. Results for $G_c$ and $G$ for the
largest field, $h=0.6$ are shown in figure~\ref{SK_h06_G} for sizes
$N=7,11,13,23,33$.  Results for $ A_c$, the Binder ratio and its
corresponding connected quantity, $B_c$, are displayed in
figures~\ref{SK_h_ac},~\ref{SK_h_bind} and \ref{SK_h_bindc}, for both
applied fields, $h=0.3$ (a) and $h=0.6$ (b).

\begin{itemize}
\item{Small fields: $h=0.3$.}

First of all, let us focus our attention on the results for
$h=0.3$. In figures~\ref{SK_h03_G} and ~\ref{SK_h_ac} we observe that
$G$, $G_c$, $A$ and $A_c$ display a clear crossing of the curves which
is not seen neither for the Binder ratio where curves for all sizes
seem to merge in the low temperature (figure~\ref{SK_h_bind}~(a)), nor
for the connected Binder cumulant
\nolinebreak(figure~\ref{SK_h_bindc}(a)). Despite, there is a clear
difference between disconnected parameters ($G$ and $A$) and connected
ones ($G_c$ and $A_c$). The crossing for disconnected quantities, $A$
and $G$, takes place at a temperature which is higher than $1$ for
small samples and slowly approaches $1$ as we increase the size. On
the contrary, results for $A_c$ and $G_c$, display a crossing, $T^*$,
at a temperature which already for small samples is smaller than one,
and that approaches the predicted $T_c=0.65$ as we increase the size
of the system.  Indeed, for the largest sizes $N=29, 33,64$ curves
cross around $T=0.65$ (figure~\ref{SK_h03_G} (c)).
\begin{figure}[tbp]
\begin{center}
\includegraphics*[width=12cm,height=10cm]{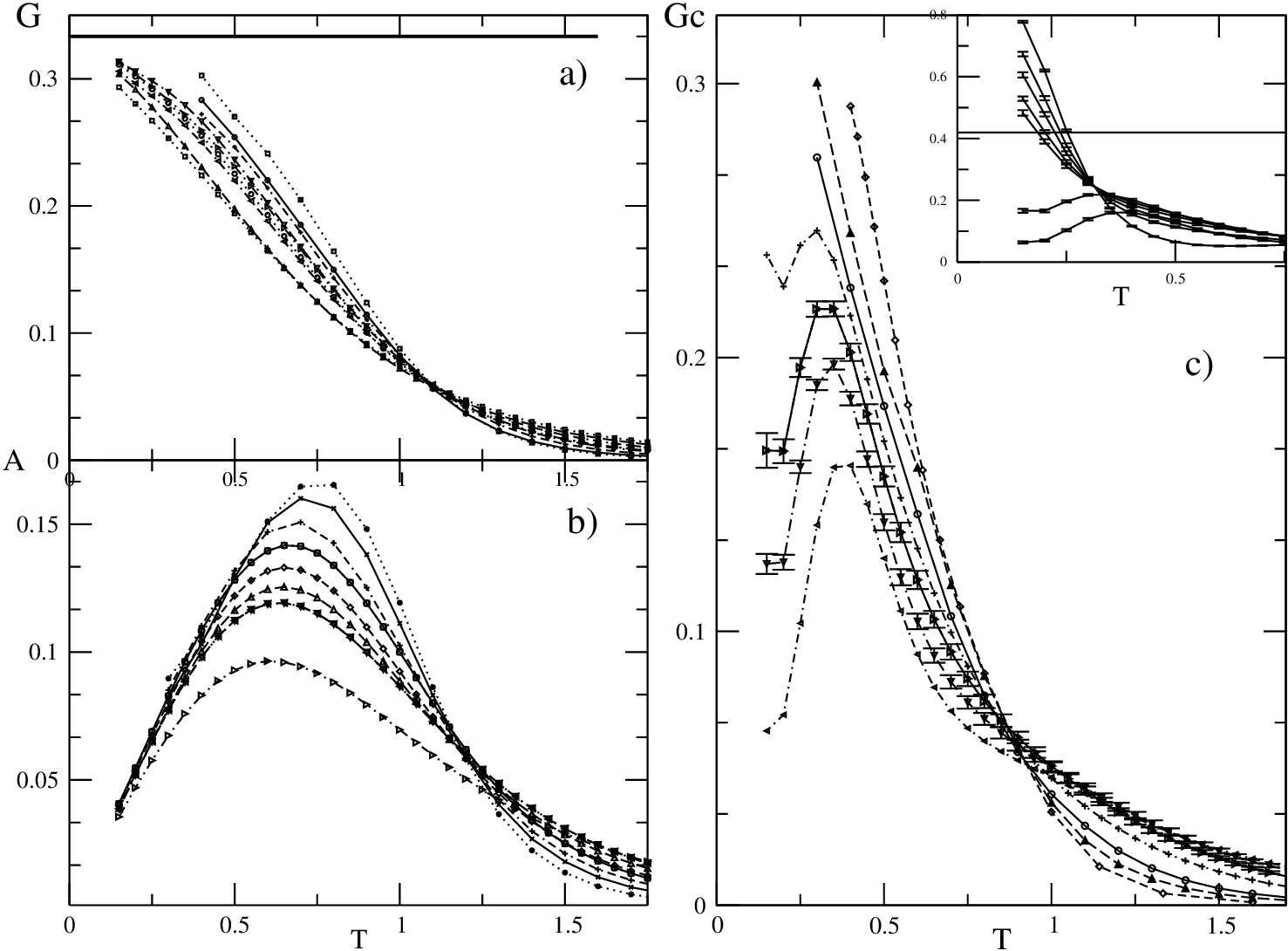}
\vskip 0.05in
\caption{Results for the $G$,$A$, and $G_c$ for the SK model in a field
$h=0.3$. a) Curves for $G$ for $N=5,6,7,9,11$-~dashed lines~-$,17,23,32,
64$ from bottom to top at low $T$; the full line corresponds to the
infinite-volume limit result $G=1/3$. b)~Results for $A$, for sizes:
$N=5,7,9,11,13,23,33$ (from bottom to top looking at the curves around
$T \simeq 0.75$). c)~Parameter $G_c$ for sizes:
$N=3,5,7,9,11,23,32,64$. (Note that in c) there's no full line standing
for the $T=0$ value of $G_c$ as it falls out of scale.) In the inset on
the top right corner, we show the low $T$ behavior of $G_c$ for small
samples with an even number of spins $N=2,4,6,8,10$ from top to bottom,
the full line corresponds to the $T=0$ value $G_c=13/31$. Curves below
the line correspond to samples with odd number of spins $N=3,5$.
\label{SK_h03_G}}
\end{center}
\end{figure}
\item{Large fields: $h=0.6$.}

Results in a larger field, $h=0.6$, for $G$, $G_c$ and $A_c$ displayed in
figures~\ref{SK_h06_G} and ~\ref{SK_h_ac} (b) show a greater difference
between $G$ and $G_c$. While the crossing for $G$ takes place at very high
temperatures (close to $2$) , curves for $G_c$ cross at a temperature which
even for small samples is close to $T_c=0.48$ and gets closer to it as we
increase the size. From the Binder cumulant (figure~\ref{SK_h_bind}~(b))
one gets no clear information. Numerical results display a crossing at high
temperature as curves for $G$ do, but the crossing point seems to be moving
towards higher temperatures, which suggests that this crossing has no real
connection with the existence of a transition. Still, curves for $B_c$
(figure~\ref{SK_h_bindc}~(b)) show an interesting result, since we observe
a crossing around $T^*=0.5$. Unfortunately, results are not very clean
because the transition takes place at a very low temperature and
thermalization is not easy. For this reason the crossing of curves for
$A_c$ in figure~\ref{SK_h_ac} (b) is blurred by the existence of the
divergence at $T=0$.  Notwithstanding this, our results, together with the
results in \cite{PR1} for the $P(q)$, which for very large sizes presented
two peaks in the positions predicted by the RSB solution, are the most
clear evidence for a transition in such a high field ($h=0.6$) which has
been reported up to now.
\begin{figure}[tbp]
\begin{center}
\includegraphics*[width=12cm,height=8cm]{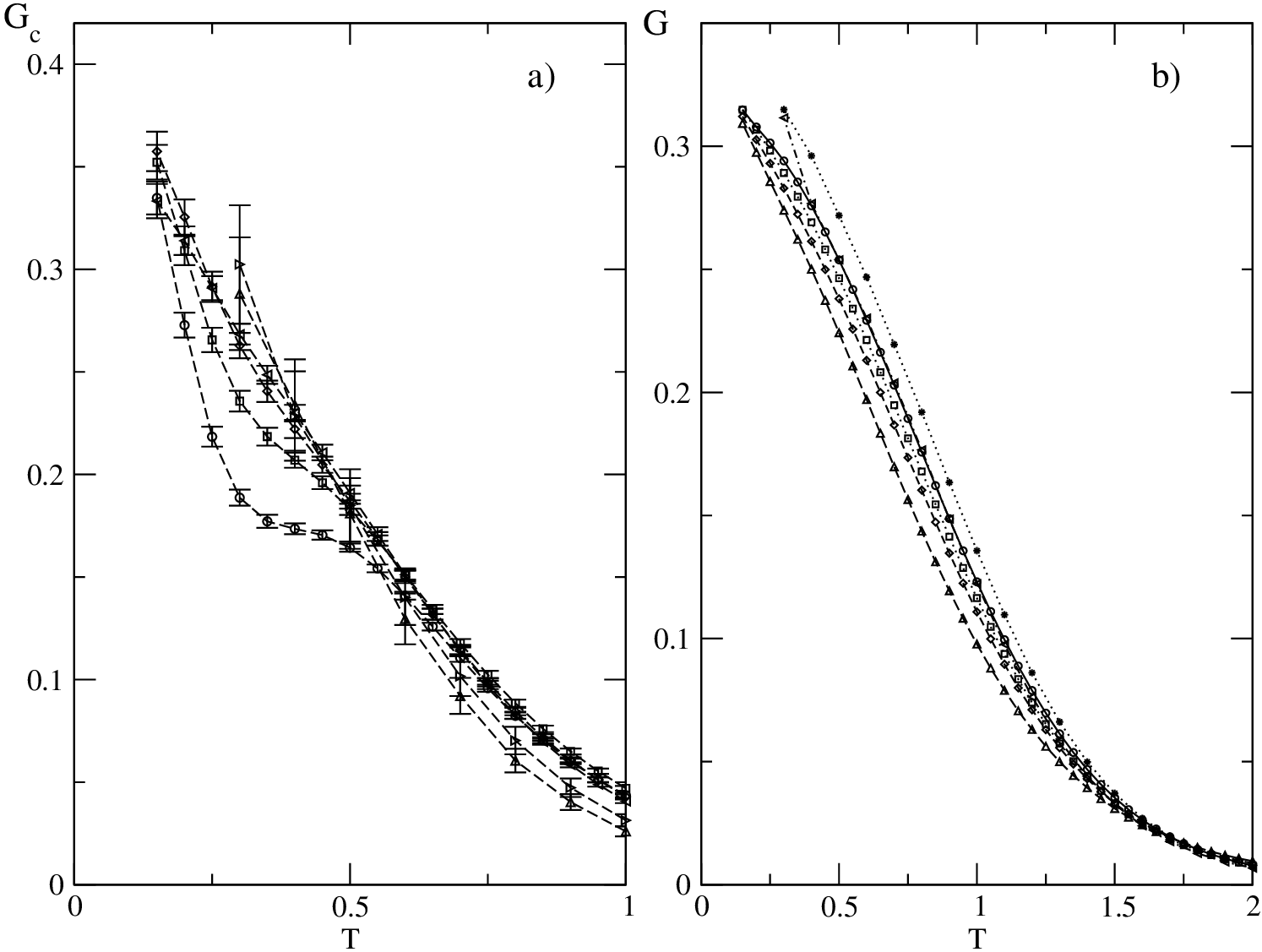}
\vskip 0.05in
\caption{$G_c$ (a) and $G$ (b) for the SK model in a field $h=0.6$ for
sizes $N=5,7,9,11,17,29$ from bottom to top in the low $T$ region.  
\label{SK_h06_G}}
\end{center}
\end{figure}

\begin{figure}
\begin{center}
\includegraphics*[width=12cm,height=8cm]{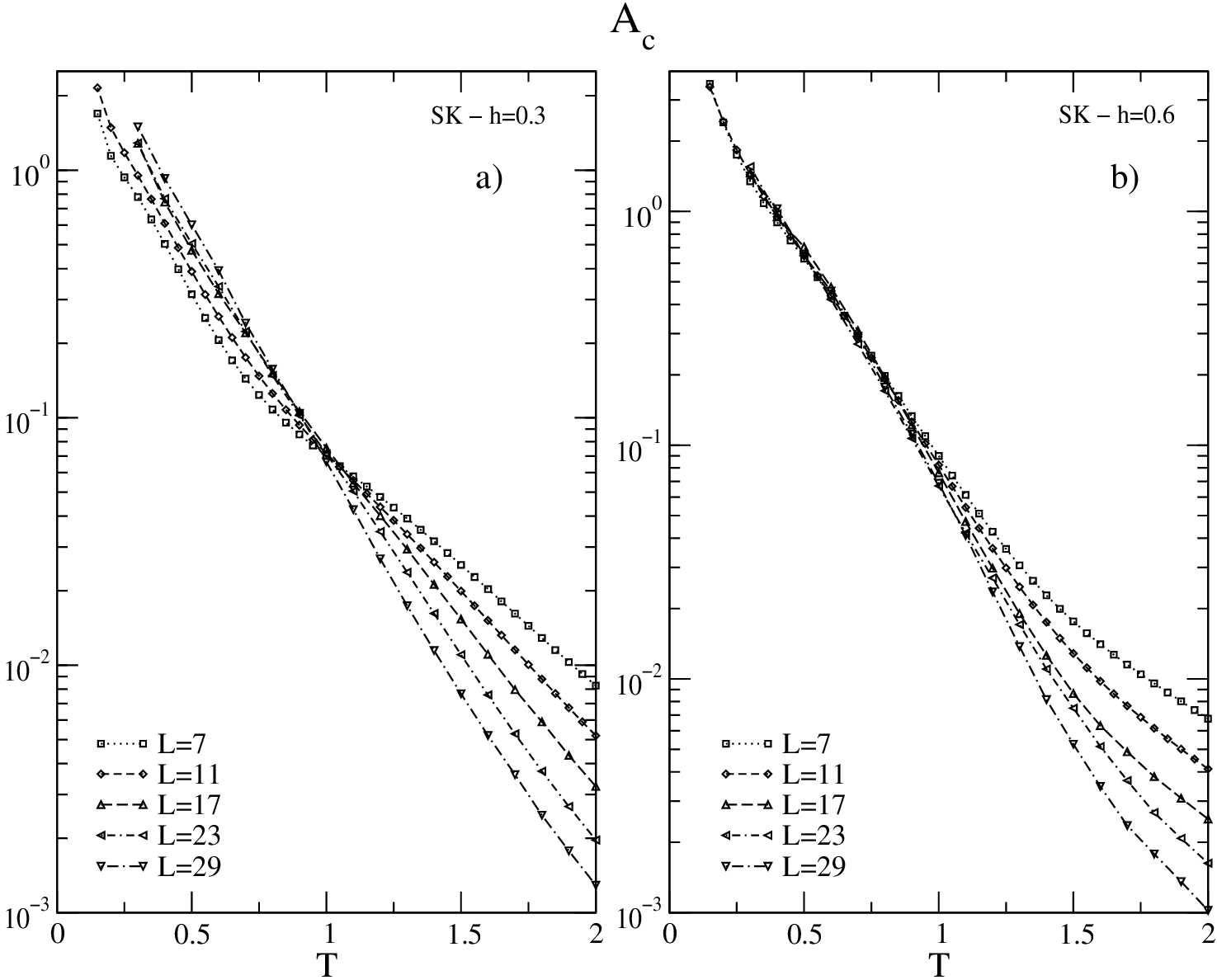}
\vskip 0.05in
\caption{  $A_c$ for the SK model in applied field for sizes
$N=7,11,13,23,29$: (a) for $h=0.3$, (b) for $h=0.6$.  
\label{SK_h_ac}}
\end{center}
\end{figure}
\begin{figure}[tbp]
\begin{center}
\includegraphics*[width=12cm,height=8cm]{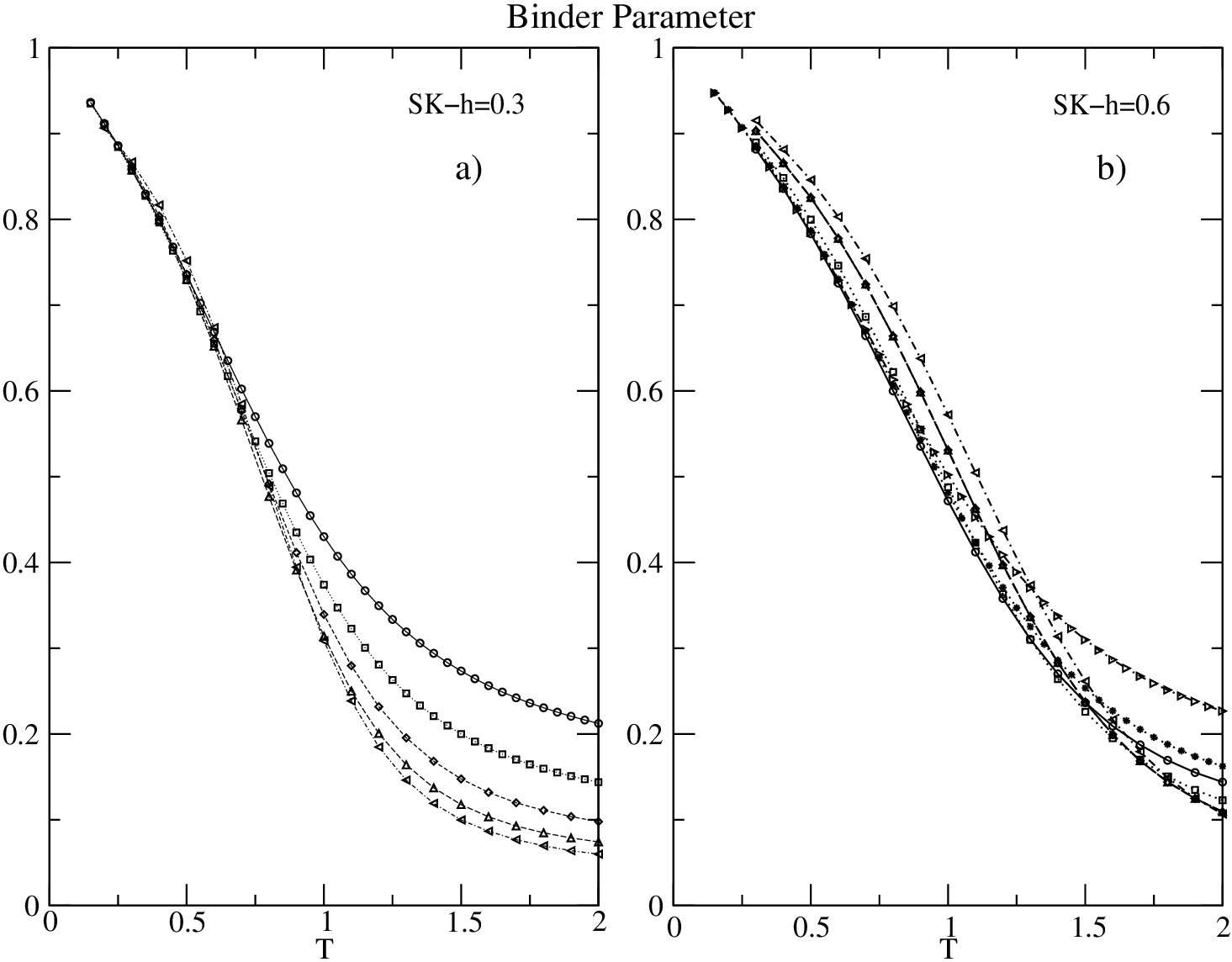}
\vskip 0.05in
\caption{Binder ratio, $B$, for the SK model in applied field for sizes
$N=7,11,13,23,33$, (a) $h=0.3$ and (b) $h=0.6$. 
\label{SK_h_bind}}
\end{center}
\end{figure}
\begin{figure}[tbp]
\begin{center}
\includegraphics*[width=12cm,height=8cm]{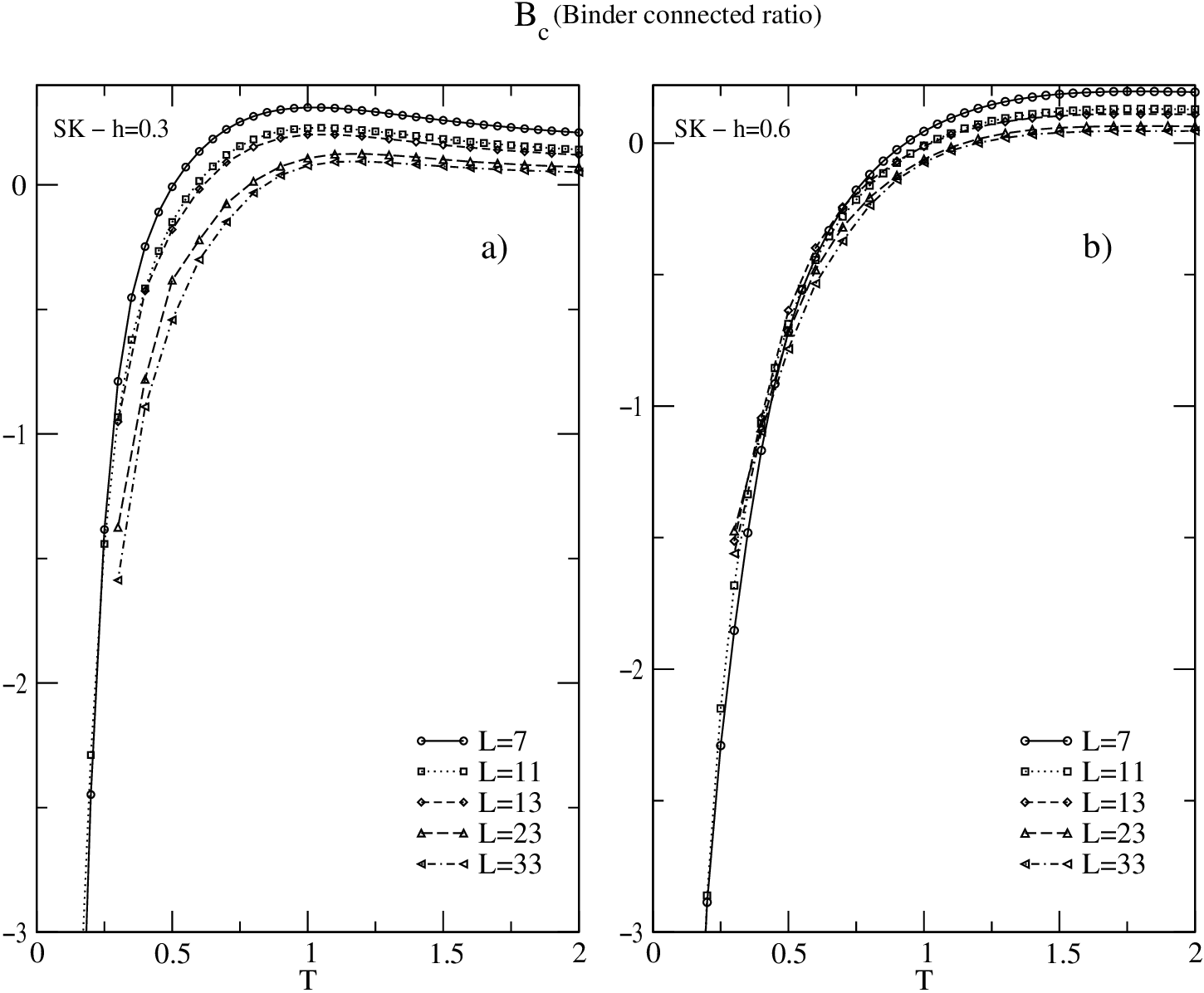}

\vskip 0.05in
\caption{Binder connected ratio, $B_c$, for the SK model in applied field
for sizes $N=7,11,13,23,33$, with $h=0.3$ on the left
and $h=0.6$ on the right.  
\label{SK_h_bindc}}
\end{center}
\end{figure}
\end{itemize}
\subsection{Summary of the results}
\begin{itemize}
\item{SK at zero field:}~ $A$ and $G$ are good parameters to locate the
transition. Finite-size effects are evidenced by a crossing at a
temperature slightly above $T_c$, that moves towards $T_c$ as size
increases. The location of the transition point is more precisely
located by $G$ and $A$ than by the Binder ratio, $B$. Finite-size
scaling analysis reveals that $B$, $G$ and $A$ have the same scaling
exponents, thus suggesting that the scaling behavior of $A$ and $G$ for
generic short-range systems may be obtained from the standard
renormalization group approach.
\item{SK in a field:}~ In general connected quantities, $G_c$, $A_c$ and
$B_c$ show a crossing which gives evidence for a transition. Finite-size
corrections to $T_c$ are of the same sign as those in the case
$h=0$. Finite-size effects are stronger at low fields because small systems
are strongly affected by the $h=0$ fixed point~\cite{FH}, suggesting that
transitions in a field need to be studied at high magnetic fields. 
\item{Continuous RSB:}~ In the frozen phase and for the largest sizes,
the general temperature behavior of the different quantities, $A$, $B$,
and $G_c$, $A_c$ and $B_c$ in models in a field, is the same with and
without a field. Therefore, it corroborates our results of the previous
section: systems exhibiting the same kind of RSB below the transition
(as in these two systems -the SK model with and without a field- which
both display full-step RSB), display qualitatively the same behavior for
$A$, $B$, $G_c$, $A_c$ and $B_c$. Note that we have intentionally
excluded $G$ from the list since it always takes the value (for
$V\to\infty$) $1/3$ in the low-temperature phase, hence does not depend
on the particular RSB pattern.
\end{itemize}

\section{Ising p-spin, p=3}
In this section we turn our attention to a model without time-reversal
symmetry where OPF are finite. We have studied the mean-field p-spin
(p=3) model with Ising spins defined by: \be {\cal H}=
-\sum_{i1,i2,i3}J_{i_1,i_2,i_3} \sigma_{i_1}\sigma_{i_2} \sigma_{i_3} \;
, \ee where couplings are Gaussian variables with zero mean and variance
$\frac{3!}{2 N^3}$. This model is well-known to have a one-step RSB
transition at $T_c\approx0.6$~ \cite{CRIS} showing a discontinuity in
the order parameter, which at the transition point jumps from zero to a
finite value. These kind of transitions are driven by a collapse of the
configurational entropy (also called entropy crisis), and have been
the object of a deep study as they have many points in common with the
glass transition in structural glasses \cite{glass}. Particularly, we
are interested in the fact that a measure of the violation of FDT in
glasses can give us information on the pattern of RSB, {\it i.e.} on the
parameter $m$ which determines the structure of the replica matrix
$Q_{ab}$.

The interest of this model relies not only on the fact that it does not
contain TRS in the Hamiltonian, but also on the fact that,  as far as OPF
do not vanish, results obtained in section III for the one-step RSB hold
for this system and, thus can provide us with a direct check of the
validity of identities (\ref{onestepgc}),~(\ref{onestepac}) and
(\ref{onestepbc}) for the finite $T$ behavior of $G_c$, $A_c$ and $B_c$
respectively. Moreover, this model presents a further simplification since
$q_0$, the off-diagonal block value of the replica matrix element $Q_{ab}$,
vanishes below the transition. For this reason, expressions
(\ref{onestepa}) and (\ref{onestepb}) given in Appendix B for $A$ and $B$
have a much simpler expression only in terms of $m\,$: 
\be
A=\frac{m}{3\,(1-m)} \; ,
\label{onestepap}
\ee 
\be
B=\frac{2-3\,m}{2\,(m-1)} \; .
\label{onestepbp}
\ee 

\begin{figure}[tbp]
\begin{center}
\includegraphics*[width=12cm,height=10cm]{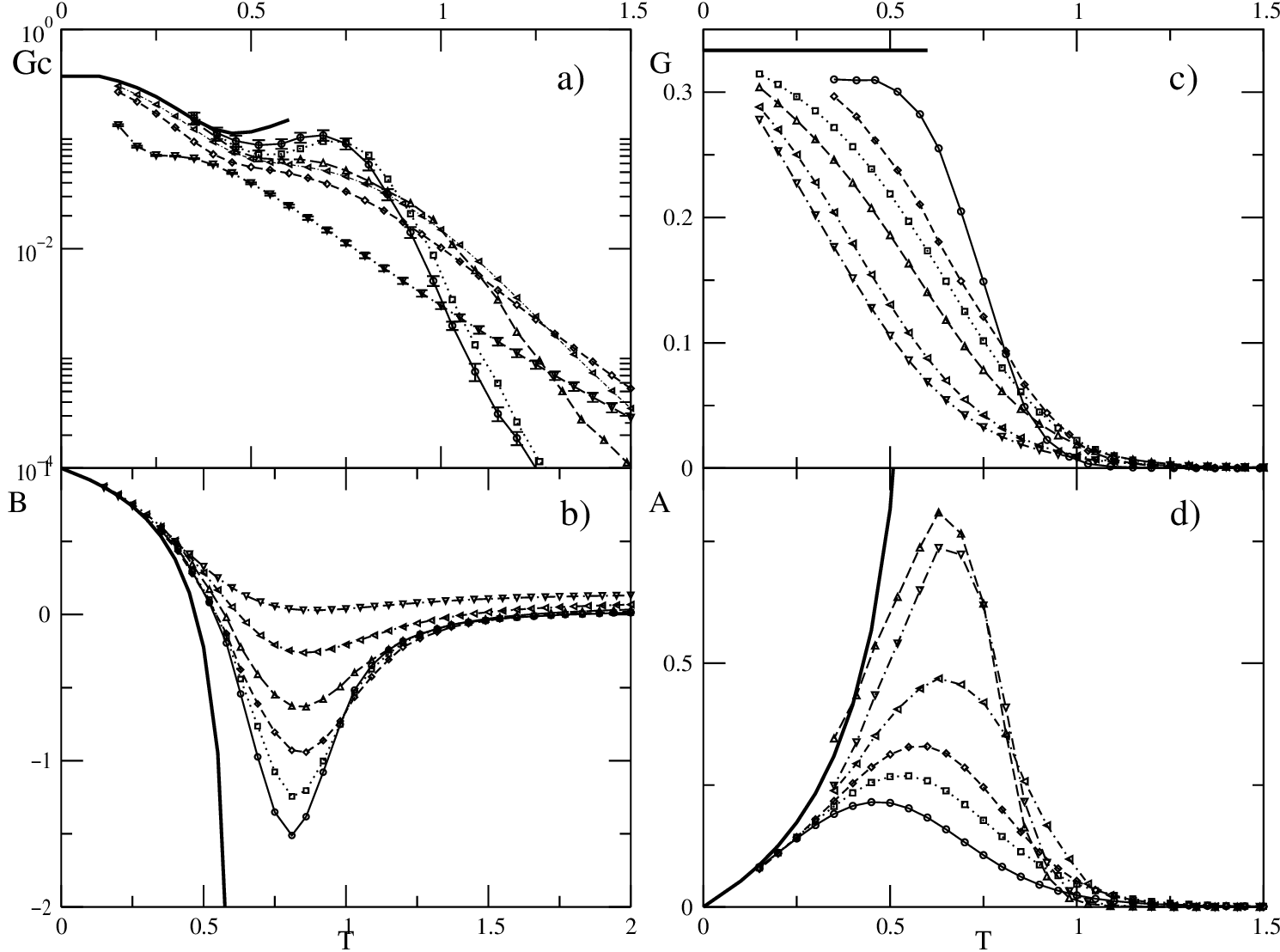}

\vskip 0.05in
\caption{We show the results for the $3$-spin model for sizes ranging from
$N=4$ to $N=33$. In every plot, the solid line corresponds to the
theoretical prediction of expressions ~(\ref{gg}), ~(\ref{onestepgc}),
~(\ref{onestepa}), and (\ref{onestepb}) for ~$G$,~$G_c$,~$A$ and ~$B_c$
respectively (see text). On the left side we display, in the top plot,
results for $G_c$ for sizes $N=4,7,11,17,29,33$ from top to bottom in the low $T$ region, errors bars are showed for
the larger and smaller sizes. Below we show the results for $B$ (Binder
ratio) for sizes $N=7,11,17,23,29,33$ from bottom to top . On the right
side, we show results for the non-connected averages, $A$ for sizes
$N=5,8,11,17,29,33$ and $G$ for sizes $N=4,8,11,17,29$ both from bottom to
top in the low-temperature region.
\label{3_spin}}
\end{center} 
\end{figure}

\begin{figure}[tbp]
\begin{center}
\includegraphics*[width=12cm,height=8cm]{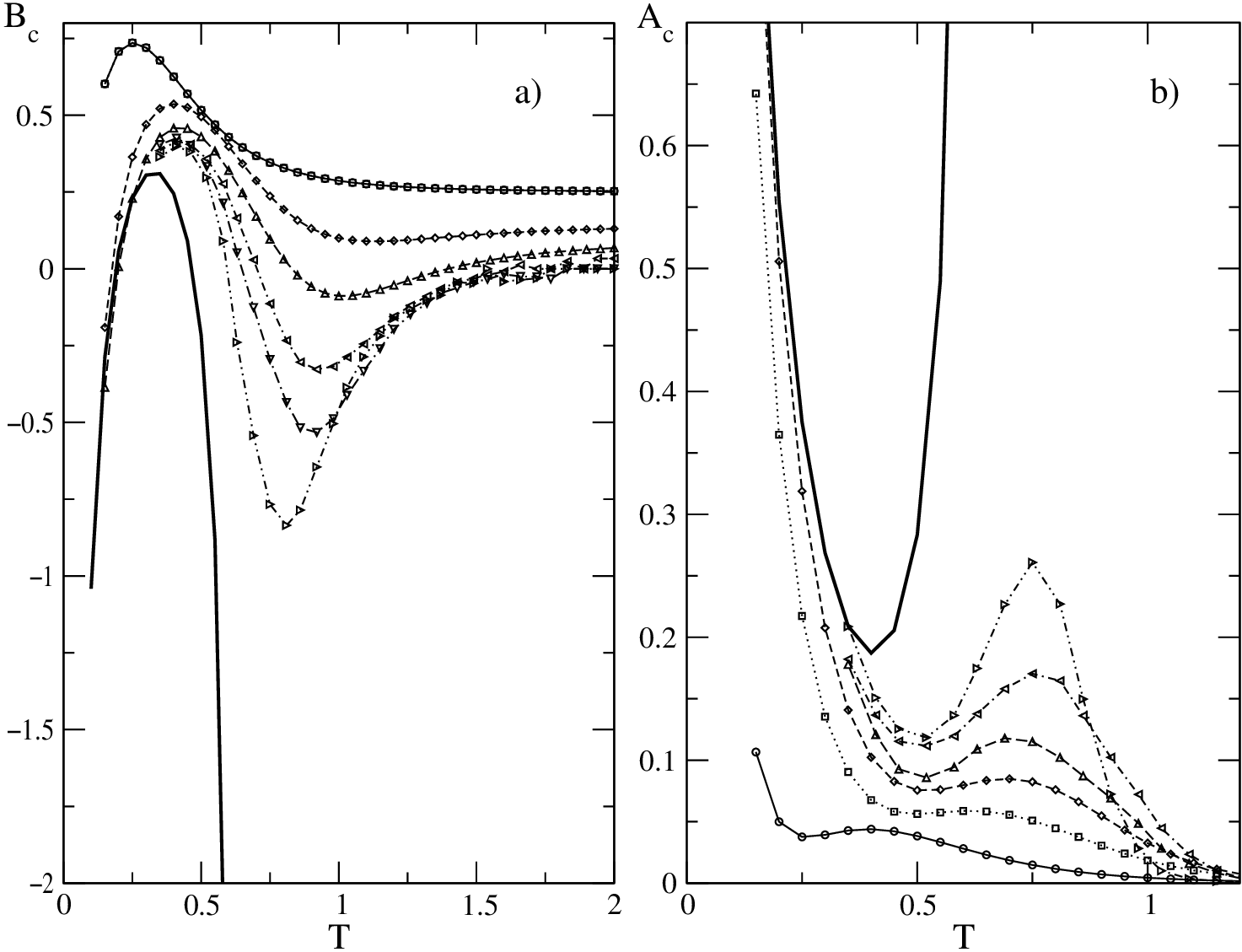}

\vskip 0.05in
\caption{Results  for the $3$-spin model: (a) $B_c$ for sizes
$N=4,7,11,17,23,29$ from top to bottom, and (b) $A_c$ for sizes
$N=4,7,11,17,23,29$ from bottom to top (in the low $T$ region). The solid line corresponds to the
numerically solved theoretical prediction for $A_c$ and $B_c$, see
eqs. (\ref{onestepac}) and (\ref{onestepbc}) respectively.  
\label{3_spin_2}}
\end{center}
\end{figure}

In order to study the behavior of OPF in this model, we have made exact
numerical computations of the partition function for sizes ranging from
$N=4$ up to $11$ averaging over $10000$ samples and Monte Carlo simulations
using the parallel tempering technique for $N=13,17,23,29,33$ averaging
over $150-500$ samples. Results for $B$, $G_c$, $G$ and $A$ are displayed
in figure ~\ref{3_spin} and for $A_c$ and $B_c$ in figure ~\ref{3_spin_2},
the full line corresponding in each case to the theoretical prediction
obtained by numerically solving the saddle point equations. We have to
point out that despite the low number of samples due to the hard task of
thermalizing these systems, results are strikingly good.

Our results for the Binder~parameter (figure~\ref{3_spin}) do not display
any crossing but show that around the transition temperature, $T_c=0.6$,
$B$ becomes negative, and has a minimum which grows together with the size
of the system as is expected from expression (\ref{onestepbp}) which
indicates that at the transition, where $m$ is set to one, the Binder ratio
has a negative divergence. Results obtained by Hukushima and Kawamura in
\cite{POTTS} for the Potts model, are different ($B$ is expected to take
the value $-1$ at the transition), since the actual values of $q_1$, $q_0$
and $m$ are different, thus corroborating our result for $B$ obtained in
Appendix B (see eq. \ref{onestepb}), which states that $B$ has a strong
dependence on the model. Results for $B_c$ are very alike, since there is
no crossing, and there is also a minimum around $T_c$ which deepens with
size. This behavior is very well understood by looking at the infinite
volume behavior of both parameters in figures \ref{3_spin} (b) and
\ref{3_spin_2} (a) in which curves for different sizes approach to the
theoretical curve from top to bottom and this is why there is no possible
crossing.

Instead, our results for $G_c$, $G$, $A$ and $A_c$ exhibit a crossing which
moves from higher to lower temperatures giving evidence for the existence
of the transition. As we had noted in the previous section, we observe that
finite-size corrections to $T^*$ are bigger in disconnected averages than
in connected ones, the formers giving a better estimate of $T_c$. However,
it is important to note that the position of the maximum of $A$ which
accounts for the existence of a phase where replica symmetry is broken has
a quick saturation towards $T=0.6$, and grows with size as is expected from
the divergence that this parameter shows at the transition
(\ref{onestepap}). A similar situation is met in results for $A_c$
(figure~\ref{3_spin_2}), since there is also a predicted divergence at
$m=1$, which is numerically observable by a maximum around $T_c$ which
grows and sharpens with size.

$ $From results for $G_c$, we can still go a bit further. We observe that, as
the size increases, the shape of the curve approaches the one of the
thermodynamic limit much faster than in the other parameters, due to the
softer behavior of $G_c$. It is interesting to note that even though we see
a crossing of the curves which moves toward $T_c$ from higher temperature
as the number of spins increases, much more clear evidence for the
existence of the transition and the kind of transition ({\it i.e.} one-step
RSB) come from the existence of a bump which becomes closer to $T_c$ as the
size grows. We observe that at low-temperatures curves always remain under
the infinite-volume line, so that there is no crossing in the low $T$
phase. Moreover, curves quickly stuck to the value of the minimum displayed
by the thermodynamic value of $G_c$. As we noted in the previous section,
the fact that this value is universal can yield valuable information about
the low-temperature phase of systems exhibiting one-step RSB.

Results in the former section showed that connected quantities, $G_c$,
$A_c$ and $B_c$ for the SK model (figures \ref{SK_h03_G}, ~\ref{SK_h06_G},
~\ref{SK_h_ac}, ~\ref{SK_h_bindc}) exhibited quite a different behavior, as
the curves were smooth and did not show any concavity, nor divergence at
the transition temperature. Hence, in addition to showing what kind of
information we can extract from the behavior of $G_c$, $A_c$, $B_c$, $A$
and $B$, we also give clear quantitative evidence of how different RSB
transitions look like extending the previous comparative analysis made in
~\cite{POTTS} between the 3-Potts model and the SK with no field.


\section{Conclusions}
In this paper we have presented a detailed analysis of parameters measuring
OPF, $A, A_c, G$ and $G_c$, as well as $B$ and $B_c$, for small
systems. Despite the numerical effort that computation of these parameters
may imply, we have shown  the powerfulness of these parameters and
what kind of information we can extract from them. Our analysis through
replica equivalence in section III, shows that the  intimate connection of
OPF with RSB yields a dependence of $A$, $B$, $G_c$, $A_c$ and $B_c$ on the
RSB pattern, in opposition to $G$, which in the SG phase takes the value
$1/3$ at $T<T_c$ ( even in marginal phases  where OPF vanish, as stated in
RS). Thus,  $A$, $B$, $G_c$, $A_c$ distinguish  between  systems displaying
different kinds of RSB in its frozen phase. Moreover, we have established a
direct connection with glasses, since we have shown that at any temperature
below $T_c$, connected quantities  depend exclusively on $m$ (which
determines the structure of the replica matrix $Q_{ab}$),
 which is the parameter having the physical meaning of an effective 
 temperature which controls the violation of FDT in structural glasses.

We have analyzed mean-field models which exhibit time-reversal symmetry in
their Hamiltonian and models which do not have this symmetry. The
universality of $G$ and $G_c$ at zero temperature stated in RS has been
checked and has been extended to the rest of quantities under study $A$,
$A_c$ and $B_c$ and we have provided with a meaningful example of when
identities~(\ref{eq5a}) and (\ref{eq5b}) do not hold, {\it i.e.} the SK model
in a field for small samples with an even number of spins. We have given
many evidences that these parameters are useful tools to locate phase
transitions into the spin-glass phase in any case. We can enumerate the
main conclusions,

\begin{enumerate}

\item{Advantage of $A,G$ (OPF) compared to $B$ (non OPF).} For models where an
analysis of the Binder cumulant is enough to show the existence of a
transition, {\it i.e.} the SK model without a field addressed in section
IV, we have shown that already for very small samples, $G$ and $A$ give
evidence for the existence of this transition. In other words,
parameters measuring OPF have smaller finite-size corrections, so that
we can extract useful information with no need of going to large
systems. Moreover we have shown that $A$ and $G$ verify scaling
relations and that the exponents coincide with those of the Binder
ratio. Obviously, $A,G$ require more computational effort than $B$ but
this effort compensates for the smallness of the corrections.

\item{Advantage of connected parameters compared to disconnected
ones.}  For models with TRS only $A,G,B$ are feasible to be
computed. The interest of parameters involving connected averages
concerns models which do not display TRS such as the SK model in a
field and the Ising 3-spin model. For those cases we have once more
checked the usefulness of these parameters to show the existence of a
transition. A comparative analysis of $G$ and $G_c$ has shown that
finite-size corrections are smaller for the latter, thus suggesting
that connected parameters such as $G_c$, $A_c$ and $B_c$ are the best
to look at in situations where there is no time-reversal symmetry. Our
results for the SK model in a field show that the effects of the $h=0$
fixed point on small sizes are very strong (in agreement with the
conventional wisdom \cite{FH,DBM}), indicating the necessity to go to
low temperatures and high fields to show the existence of a spin-glass
transition in a field. In view of some recent numerical results
\cite{HG,KPY} this goal is probably not out of reach.

\item{Temperature dependence of OPF parameters on the RSB pattern.} By
studying these mean-field models which have different kinds of RSB
transitions, we have checked that in systems where OPF do not vanish
({\it i.e.} when the results obtained through replica equivalence analysis are
meaningful), $G_c$, $A_c$, $B_c$, $A$ and $G$ have a different behavior
in the low-temperature phase. In addition, we have shown that in the
3-spin model, which has a discontinuous transition, already for very
small samples, $G_c$ yields information about the infinite-volume
behavior obtained in section III (eq.~\ref{onestepgc}) and thus about
the replica symmetry breaking pattern through the temperature dependence
of the parameter $m$.

\end{enumerate}

To sum up, our results for $G$, $G_c$, $A$ and $A_c$ for small systems, in
addition to what had already been stated in RS, {\it i.e.} the universal
values at $T=0 $ for both $G$ and $G_c$, and the trivial behavior of $G$
for any system having a SG transition, establish that parameters measuring
OPF meet all the requirements that a proper parameter to locate phase
transitions should have in disordered systems. We have studied in detail
their behavior in systems where OPF do not vanish and have shown what kind
of information small samples give about the infinite-volume behavior at
finite temperature.
In conclusion, the study of parameters measuring OPF opens a new
direction in the study of the SG phase, as they are useful not only to
locate the transition, but also to establish the nature of
spin-glasses at finite temperature. Moreover, the scaling analysis
reveals that there might be a connection between OPF and critical
fluctuations that still lacks theoretical insight. The understanding
of the trivial behavior of $G$ in the infinite-volume limit can give
us more information about the mechanisms that control this frozen SG
phase and can help us to understand the connection between
spin-glasses having a marginally stable replica symmetric phase and
systems having a broken SG phase. The further extension of these ideas
to {\em non-disordered} glass models is also an interesting open
problem.

\vskip 1cm 

{\bf Acknowledgments:} F.~R. is supported by the Ministerio
de Educaci\'on y Ciencia in Spain (PB97-0971). M.~S. is supported by the
Ministerio de Educaci\'on y Ciencia of Spain, grant AP-98
36523875. M.~P. and F.~R. acknowledge financial support by the
French-Spanish Picasso program (Acciones Integradas HF1998-0097).

\vskip 0.2in
\appendix
\begin{center}
{\bf Appendix A: Low-temperature behavior of $A$,
$G$, $B$ and $A_c$, $G_c$, $B_c$.}
\end{center}
\vskip 0.1in
In this Appendix we give a brief summary of the main points of the proof of
the conjecture presented in RS, and show how the zero temperature values of
all the parameters introduced in section II are obtained. We consider the
most general finite spin system of Ising variables, described by the local
field Hamiltonian:  
\be
{\cal H}=-\sum_{i}h_i\s_i \; ,
\label{eqa1}
\ee
under two hypothesis: 
\begin{itemize}
\item it has a unique ground state (up to a global flip of all the spins),
given by the configuration $\{\s^*\}$ : $E_0=-\sum_i h_i^*
\s_i^*$. 
\item its local field probability distribution has a finite weight at zero
field. 
\end{itemize}
Under these two assumptions it was shown in RS that the low-temperature
behavior of all the terms which appear in the definition of $G$ and $G_c$
(and hence in all the other quantities defined in section II) are dominated
by one spin excitations \cite{FOOTNOTE}. This is because one-spin
excitations yield a linear contribution in $T$, while two spin excitations
yield a contribution of order $(T^2)$. Thus, in any general {\em finite}
system where all excitations are possible, one-spin excitations give the
dominant contribution at low $T$ and are responsible for the universal
values $G=\frac{1}{3}$ and $G_c=\frac{13}{31}$ at $T=0$. In cases where
the lowest excitation systematically involves clusters 
of spins a generalization of the present argument shows that the present
results still hold \cite{RS2}.

The argument proceeds as follows. First of all we note that all elements
appearing in the definition of $G$ and $A$ can be expressed in terms of the
two and four point correlation functions. If we define
$T_{ij}=\lan\s_i\s_j\ran^2$ and $T_{ijkl}=\lan\s_i\s_j\s_k\s_l\ran^2$, then
we have: 
\be
\langle q^2\rangle = \frac{1}{V}+\frac{1}{V^2}\sum_{i\ne
j}T_{ij} \; ,
\label{eqa2}
\ee
\be
\langle q^2\rangle^2=\frac{1}{V^2}+\frac{2}{V^3}\sum_{i\ne j}T_{ij}
+\frac{2}{V^4}\sum_{i\ne j}T_{ij}^2+\frac{4}{V^4}\sum_{(i\ne j \ne
k)}T_{ij}T_{ik} +\frac{1}{V^4}\sum_{(i\ne j\ne k\ne l)}T_{ij}T_{kl} \; ,
\label{eqa4}
\ee
\be
\langle q^4\rangle=\frac{1}{V^4}\Bigl ( 3V^2-2 V +
(6V-8)\sum_{i\ne j}T_{ij} +\sum_{(i,j,k,l)}T_{ijkl}\Bigr ) \; .
\label{eqa5}
\ee
Terms appearing in the definitions of the connected quantities also depend
on one and three points correlation functions. With the definitions
$T_{i}=\lan\s_i\ran^2$ and $T_{ijk}=\lan\s_i\s_j\s_k\ran^2$, we
obtain the following expressions :

\be 
\langle q\rangle^4=\frac{1}{V^4}\Bigl ( \sum _i T_i^4+ 3
\sum_{i\ne j}T_{i}^2T_{j}^2 +4 \sum_{i\ne j}T_i^3 T_j+6 \sum_{(i,j,k)}T_i^2
T_j T_k\Bigr ) \; ,
\label{eqa5b}
\ee
\be  
\langle q^3\rangle\langle q\rangle=\frac{1}{V^4}\Bigl ((3 V-2) \sum _i
T_i^2+ (3V-2) \sum_{i\ne j}T_{i}^2T_{j}^2 +3 \sum_{(i,j,k)}T_i T_{ijk}+
\sum_{(i,j,k,l)}T_{i,j,k}T_l\Bigr )\; ,
\label{eqa6}
\ee
\be  
\langle q^2\rangle\langle q\rangle^2=\frac{1}{V^4}\Bigl ( V \sum _i T_i^2+
V \sum_{i\ne j}T_{i}T_{j} +\sum_{(i,j,k)}T_i^2 T_{jk}+ 4\sum_{i\ne
j}T_{i}T_{j}T_{ij} +2\sum_{i\ne j}T_i^2 T_{ij}+\sum_{(i,j,k,l)} T_{i,j}T_k
T_l\Bigr ) \; . 
\label{eqa7}
\ee
{\bf  One-spin excitations.}\\
\nopagebreak
We consider now all possible one-spin excitations. The energy cost of
reversing the spin $i$ will be: $e_i= 2 h_i^* \s_i^*$. From here one can
compute the correlation functions previously introduced and obtain:

\be
\lan \s_i\ran=\s_i^*\left(1-\frac{2 \exp(-\beta e_i)}{1+\sum_{l=1,v}\exp(-2
\beta e_l)}\right)\approx \s_i^*\left(1-2 x_i\right) \; , 
\label{eqa8}
\ee
\be
\lan \s_i\s_j\ran\approx \s_i^*\s_j^*\left(1-2 (x_i+ x_j)\right) \; ,
\label{eqa9}
\ee
\be
\lan \s_i\s_j\s_k\ran\approx \s_i^*\s_j^*\s_k^*\left(1-2 (x_i+
x_j+x_k)\right) \; ,
\label{eqa10}
\ee
\be
\lan \s_i\s_j\s_k\s_l\ran\approx \s_i^*\s_j^*\s_k^*\s_l^*\left(1-2 (x_i+
x_j+x_k+x_l)\right) \; ,
\label{eqa11}
\ee 
with the definition $x_i\equiv\exp(-\beta e_i)$. Note that we have made
the approximation $x_i<<1$, thus we can approximate the denominator in
(\ref{eqa8}) by one. We have to stress that, as widely discussed in RS, this
approximation implies that we are in the limit $\beta V <<1$. This is a key
point in our argument concerning the local field probability distribution. 

When performing the averages over the disorder one has to tackle with the
following objects: 
\be
\overline{x_i^mx_j^n}\equiv \int dh_i\, dh_j\,
\hat{P}(h_i,h_j)\,e^{-2\beta( m\; h_i^* \s_i^*+ n\; h_j^* \s_j^*)} 
\label{eqa12}
\ee
where the probability distribution of two fields results from integrating
out the rest of the fields: 
\be
 \hat{P}(h^*_i,h^*_j)\equiv\int \prod_{k\ne i,j}\,dh^*_k\, P(h^*_1,....,h^*_V)~~~.
\label{eqa13}
\ee
We have to point out that two local fields may be correlated so that the
probability distribution of one field $\hat{P}(h^*_i)\equiv P_i$ depends on
the site. However, the probability of having two equal local fields is very
small ~\cite{FOOTNOTE}, thus in (\ref{eqa13}) there are no contributions of
the type $\delta(h^*_i-h^*_j)$. It then follows from expression
(\ref{eqa12}) by a simple saddle-point calculation that terms which
depend on two sites have a contribution $\propto {\cal O} (T^2)$ while terms
which depend only on one site have contributions $\propto {\cal O} (T)$ :

\be
\overline{x_i^mx_j^n}=\frac{T\, P_{ij}(0,0)}{m\; n}\; \;\;\;\;\; 
\overline{x_i^m}=\frac{T P_i(0)}{m}~~~. 
\label{eq14a}
\ee
By looking at the above expressions for the correlation functions one
observes that terms which depend on only one site are present only in
one-spin excitations since, as shown in RS, correlation functions
computed in the case of two-spin excitations depend on the products
$x_ix_j$, thus bringing sub dominant contributions.

Then, after some algebra, the results obtained for the quantities appearing
in the definitions of $G$, $G_c$, $A$, $A_c$ and $B$ are the following:  
\begin{eqnarray}
\overline{\langle q^2\rangle^2}-\overline{\langle q^2\rangle}^2 &=& 
\frac{16\, T \sum_i P_i(0)}{3V^4}(V-1)^2\,+\,{\cal O}(T^2) \; ,
\label{eqa15}
\\
\overline{\langle q^4\rangle}-\overline{\langle q^2\rangle}^2 &=& 
\frac{16 \,T \sum_i P_i(0)}{V^4}(V-1)^2\,+\,{\cal O}(T^2) \; ,
\label{eqa16}
\\
\overline{\langle (q-\langle q\rangle)^2\rangle^2}&=&\frac{218\,T \sum_i
P_i(0)}{105\,V^4}+\; {\cal O}(T^2) \; , 
\label{eqa17}
\\
\overline{\langle (q-\langle q\rangle)^2\rangle}&=&\frac{8 \,T \sum_i
P_i(0)}{3\, V^4}+\,{\cal O}(T^2) \; , 
\label{eqa18}
\\
\overline{\langle (q-\langle
q\rangle)^4\rangle}&=&\frac{1}{V^4}\frac{496\,T\sum_i P_i(0)}{105\,V^4}+\;
{\cal O}(T^2) \; . 
\label{eqa19}
\end{eqnarray}
Then we obtain for $G$ and $G_c$ for any finite volume, 
\be
G=\frac{1}{3}+ {\cal O}(T) \;~~~~ G_c=\frac{13}{31} + {\cal O}(T) \; .
\label{eqa20}
\ee

The situation is rather different for $A_c$ and $B_c$. From the expression
of eq.~(\ref{eq4}) of $A_c$, we see that the numerator is linear in $T$,
while the denominator $\overline{\langle (q-\langle q\rangle)^2\rangle}^2 $
is quadratic in $T$. Hence we have that in the low-temperature limit: 
\be
\lim_{T\to 0}\; A_c \propto \frac{1}{T}~\rightarrow~\infty \; .
\label{eqa21}
\ee
The same situation occurs for the computation of $B_c$ given by
eq.~(\ref{eq4bb}) where we have to compare $\overline{\langle (q-\langle 
q\rangle)^4\rangle} \propto T $ with   $\overline{\langle (q-\langle
q\rangle)^2\rangle}^2 \propto T^2$. It then follows that: 
\be
\lim_{T\to 0}\; B_c~\propto \,-\,\frac{1}{T}~\rightarrow\,-\infty \; .
\label{eqa22}
\ee

\vskip 0.2in   
\appendix
\begin{center}
{\bf Appendix B: Using replica equivalence to compute OPF.}\\
\end{center}
\vskip 0.1in
In this appendix we report  explicit calculations of some results outlined
in Section III. In particular, we describe in detail how  to make use of
the replica equivalence property which states that quantities such as
$\sum_{a}Q_{ab}^k$ do not depend on the replica index $b$, to compute the
different terms involved in $G$ and $G_c$ which lead to identities
(\ref{eq5b}) and (\ref{onestepgc}) respectively. 

In general, we have to deal with terms involving one or more overlaps which
in terms of the replica matrix elements $Q_{ab}$ read :
\begin{eqnarray}
\overline{\lan q^k \ran}=\frac{1}{n(n-1)}\sum_{ab}Q_{ab}^k &=& 
\int dq P(q) q^k\equiv q_k  \; ,
\label {eq1b}\\
\overline{\lan q
\ran^y}=\frac{\sum_{a,b,c,d...,l_{2(y-1)},l_{2y-1}}
Q_{ab}Q_{cd}..Q_{l_{2y-1},l_{2y}}}{n(n-1)(n-2)...(n-(2y-1))}
&=& \int dq_1dq_2.....dq_y P(q_1,q_2....q_y) q_1q_2.....q_y \; ,  
\label {eq2b}
\end{eqnarray}
where sums always run over different indexes.   Note that we have defined
$q_k$ in order to simplify the notation. As we have already said in section
III this can also be expressed in terms of the $P(q)$ for a different
number of replicas as we show in the last equality. \\ 
\vskip 0.1in
{\bf Computation of $G$}\\
\vskip 0.1in
The terms which appear in the definition of $G$ are the following:
 \begin{eqnarray}
\overline{\lan q^2 \ran}&=& \frac{1}{n(n-1)}\sum_{ab}Q_{ab}^2
=\int dq P(q) q^2\equiv q_2 \; , 
\label {eqq2b}\\
\overline{\lan q^4 \ran}&=&\frac{1}{n(n-1)}\sum_{ab}Q_{ab}^4=\int dq P(q)
q^4\equiv q_4  \; ,
\label {eqq4b}\\
\overline{\lan q^2 \ran^2}&=&\frac{1}{n(n-1)(n-2)(n-3)}
\sum_{a,b,c,d}Q_{ab}^2Q_{cd}^2=\int  dqdq_1 P(q,q_1) q^2q_1^2 \; .
\label {eqq2q2b}
\end{eqnarray}
Thus we only have to deal with terms involving one and two overlaps.
As is shown in \cite{MPSV} and discussed in detail in  \cite{PARISI2}, the
RE property allows to express the probability distribution of two overlaps
in terms of the probability distribution of one overlap: $P(q)$. Hence
$\overline{\lan q^2 \ran^2}$ can be expressed in terms of $\overline{\lan
q^4 \ran}$  $\overline{\lan q^2 \ran}$ and $\overline{\lan q \ran}$. 
\begin{itemize}
\item Quantities involving two overlaps.\\
The general term reads :
\be
\overline{\lan q^k \ran \lan q^l
\ran}=\frac{1}{n(n-1)(n-2)(n-3)}\sum_{a,b,c,d}Q_{ab}^kQ_{cd}^l \; .
\label{qkql}
\ee 
The calculation proceeds as follows: the sum appearing in (\ref{qkql})
can be re-expressed as: 
\be
\sum_{a,b,c,d}Q_{ab}^kQ_{cd}^l = \sum_{ab}Q_{ab}^k\sum_{ab} Q_{ab}^l
-2\sum_{ab}Q_{ab}^{k+l}-4\sum_{a,b,c}Q_{ab}^kQ_{ac}^l \; .    
\label{eq3b}
\ee 
In order to compute the last term of the previous identity we use replica
equivalence, {\it i.e.} : 
\be
\sum_{c}Q_{ac}=\frac{1}{n}\sum_{ac}Q_{ac} \; . 
\label{eq4b}
\ee
Thus, for the last term in  (\ref{eq3b}), we obtain:
\be
\sum_{a,b,c}Q_{ab}^kQ_{ac}^l = \sum_{ab}Q_{ab}^k\sum_{c} Q_{ac}^l =
\sum_{ab}Q_{ab}^k\left(\frac{\sum_{ac}Q_{ac}^l}{n}-Q_{ab}^l\right) \; , 
\label{eq5cb}
\ee
so that substituting this last result into identity (\ref{eqq2q2b}) we
obtain the following general relation: 
\begin{eqnarray}
\overline{\lan q^k \ran \lan q^l \ran}
&=&
\frac{(n-4)(n-1)}{(n-2)(n-3)}\frac{\sum_{ab}Q_{ab}^k}{n(n-1)}
\frac{\sum_{ab}Q_{ab}^l}{n(n-1)} +\frac{2}{(n-2)(n-3)}
\frac{\sum_{ab}Q_{ab}^{k+l}}{n(n-1)}\nn \\
&=& \frac{(n-4)(n-1)}{(n-2)(n-3)} q_k q_l+\frac{2}{(n-2)(n-3)} q_{k+l}\; .  
\label{eq7b}
\end{eqnarray}
>From this relation, the computation of $\overline{\lan q^2 \ran^2}$ and 
terms such as $\overline{\lan q \ran\ \lan q^3 \ran}$ and  $\overline{\lan
q \ran^2}$ which appear in the definition of $G_c$ is straightforward. Thus
for $\overline{\lan q^2 \ran^2}$ we obtain:  
\bea
\overline{\lan q^2 \ran \lan q^2 \ran}
&=&\frac{(n-4)(n-1)}{(n-2)(n-3)}
\left(\frac{\sum_{ab}Q_{ab}^2}{n(n-1)}\right)^2+\frac{2}{(n-2)(n-3)} 
\frac{\sum_{ab}Q_{ab}^{4}}{n(n-1)}\nn \\
&=&\frac{(n-4)(n-1)}{(n-2)(n-3)} 
q_2^2+\frac{2}{(n-2)(n-3)} q_{4} \; . 
\label{eq6b}
\eea
\end{itemize}
At this point we are able to compute the value of the numerator and
denominator of $G$, which yield the well-known result: 
\bea
\hbox{numerator} &=&\frac{2}{(n-2)(n-3)}(\int dq P(q) q^4-(\int dq P(q)
q^2)^2)\nn\\ 
\hbox{denominator} &=& \int dq P(q) q^4-(\int dq P(q) q^2)^2 \nn \\
&& \rightarrow \lim_{n\to 0}~~G =
\frac{numerator}{denominator}=\frac{1}{3} \; ,  
\label{eq8b}
\eea
provided OPF $(\int dq P(q) q^4-(\int dq P(q) q^2)^2)$ do not vanish. For
the parameter $A$ this yields in the limit $n\to 0$ : 
\be
A=\frac{q_4-q_2^2}{3\;q_2^2} \; .
\label{eq8bb}
\ee 
\vskip 0.1in
{\bf Computation of $G_c$}\\
\vskip 0.1in
Now we have to deal with harder terms which appear in the calculation of
$G_c$ and involve the probability distribution of three and four
overlaps. The numerator reads: $\overline{\lan q^2 \ran^2}-2\overline{\lan
q^2 \ran\lan q \ran^2}+\overline{\lan q \ran^4}-(\overline{\lan q^2
\ran}-\overline{\lan q \ran^2})^2$ and the denominator reads:
$\overline{\lan q^4 \ran}+6\overline{\lan q^2 \ran\lan q
\ran^2}-3\overline{\lan q \ran^4}-4\overline{\lan q^3 \ran\lan q
\ran}-(\overline{\lan q^2 \ran}-\overline{\lan q \ran^2})^2$ so that we
have to compute: 
\begin{eqnarray}
\overline{\lan q^2 \ran\lan q
\ran^2}=\frac{1}{n(n-1)(n-2)(n-3)(n-4)(n-5)}\sum_{a,b,c,d,e,f}Q_{ab}^2Q_{cd}Q_{ef}\ 
\label{eqq2qqb}\\
\overline{\lan q
\ran^4}=\frac{1}{n(n-1)(n-2)(n-3)(n-4)(n-5)(n-6)(n-7)}
\sum_{a,b,c,d,e,f,g,h}Q_{ab}Q_{cd}Q_{ef}Q_{gh} \; . 
\label{eqqqqqb}
\end{eqnarray}
As we have already pointed in section III, these terms will depend in
general in terms containing $P(q)$ and terms which depend on the
probability distribution of $3$ or $4$ connected overlaps, terms
like $\sum_{a,b,c}Q_{ab}^kQ_{bc}^lQ_{ca}^p$ and
$\sum_{a,b,c,d}Q_{ab}^kQ_{bc}^lQ_{ca}^pQ_{ca}^t$ where replica equivalence
cannot be used since we have in general that: $\sum_{b}Q_{ab}^kQ_{bc}^l$
depends on the replica indices $a$ and $c$, so that for any particular
structure of the replica matrix $ Q_{ab}$ will have different values.
\begin{itemize}
\item Quantities involving three overlaps.\\
Now we will derive the general relation for any term containing three
overlaps such as (\ref{eqq2qqb}). The general term is: 
\be
\sum_{a,b,c,d,e,f}Q_{ab}^kQ_{cd}^lQ_{ef}^m = 
\sum_{a,b,c,d}Q_{ab}^kQ_{cd}^l\sum_{e,f}Q_{ef}^m-4
\sum_{a,b,c,d,e}Q_{ab}^kQ_{cd}^lQ_{eb}^m\;- 4
\sum_{a,b,c,d,e}Q_{ab}^kQ_{cd}^lQ_{ec}^m \; . 
\label{eq10b} 
\ee
The first term can be computed from identity (\ref{eq7b}), since we have: 
\be
\frac{1}{n(n-1)(n-2)(n-3)(n-4)(n-5)}
\sum_{(a,b,c,d}Q_{ab}^kQ_{cd}^l\sum_{e,f}Q_{ef}^m =
\frac{n(n-1)}{(n-4)(n-5)}\overline{\lan 
q^{l} \ran\lan q^{k} \ran}\overline{\lan q^{m} \ran} \; . 
\label{eq11b}
\ee
The other two terms on the  r.h.s of relation (\ref{eq10b}) are identical
by permuting sub-indices $(d,b)$ and super-indices $(l,k)$ so that we only
have to compute one of them: $\sum_{a,b,c,d,e}Q_{ab}^kQ_{cd}^lQ_{eb}^m$.
We can apply  RE (\ref{eq4b}) to the sum of index $e$ and obtain:  
\be
\sum_{a,b,c,d,e} Q_{ab}^k Q_{cd}^l Q_{eb}^m = \sum_{a,b,c,d} Q_{ab}^k
Q_{cd}^l\left(\frac{\sum_{e,b} Q_{eb}^m}{n}\right) - \sum_{a,b,c,d}
Q_{ab}^{k+m} Q_{cd}^l -2\sum_{a,b,c,d}Q_{ab}^{k} Q_{cd}^l Q_{cb}^m \; .  
\label{eq12b} 
\ee
Again we make use of identity (\ref{eq7b}) to compute the two first terms
on the r.h.s of the previous identity, so that we are only left with the
last term: $\sum_{a,b,c,d}Q_{ab}^{k}Q_{cd}^lQ_{cb}^m$. By applying the RE to
the sum over $d$ we finally get: 
\be
\sum_{a,b,c,d}Q_{ab}^{k}Q_{cd}^lQ_{cb}^m= \sum_{a,b,c}
Q_{ab}^{k}Q_{cb}^m\left(\frac{\sum_{d,c}Q_{d,c}^l}{n}\right)-\sum_{a,b,c}
Q_{ab}^{k}Q_{cb}^m Q_{ca}^l-\sum_{a,b,c} Q_{ab}^{k}Q_{cb}^{m+l} \; . 
\label{eq13b} 
\ee
So, now, we know how to compute all the terms in function of the
probability distribution of one overlap except for the term $\sum_{a,b,c}
Q_{ab}^{k}Q_{cb}^m Q_{ca}^l$. 
Putting all terms together we obtain the following expression for the
general element which involves the probability distribution of three
overlaps, 
\begin{eqnarray}
\label{eq14b}
\overline{\lan q^k \ran\lan q^l \ran\lan q^m \ran}&=&
\frac{(n-4)(n-8)+8(n-1))(n-1)^2}{(n-2)(n-3)(n-4)(n-5)}
\frac{\sum_{ab}Q_{ab}^k}{n(n-1)}\frac{\sum_{ab}Q_{ab}^l}{n(n-1)}
\frac{\sum_{ab}Q_{ab}^m}{n(n-1)}\\ 
&+&\frac{16\sum_{ab}Q_{ab}^{l+m+k}-8(n-2)\sum_{a,b,c}
Q_{ab}^{k}Q_{cb}^{m}Q_{ca}^{l}}{n(n-1)(n-2)(n-3)(n-4)(n-5)}\nn\\
&+& \frac{2(n-8)(n-1)}{(n-2)(n-3)(n-4)(n-5)}
(\frac{\sum_{ab}Q_{ab}^{k+l}}{n(n-1)}\frac{\sum_{ab}Q_{ab}^{m}}{n(n-1)} 
+\frac{\sum_{ab}Q_{ab}^{m+l}}{n(n-1)}\frac{\sum_{ab}Q_{ab}^{k}}{n(n-1)}\nn\\
&+&\frac{\sum_{ab}Q_{ab}^{k+m}}{n(n-1)}
\frac{\sum_{ab}Q_{ab}^{l}}{n(n-1)})\nn\\ 
&=& \frac{(n-4)(n-8)+8(n-1))(n-1)^2}{(n-2)(n-3)(n-4)(n-5)}~q_k q_l q_m \nn\\
&+& \frac{2(n-8)(n-1)}{(n-2)(n-3)(n-4)(n-5)}~(q_k q_{m+l}+q_l q_{m+k}+q_m
q_{k+l})\nn\\
&+& \frac{16}{(n-2)(n-3)(n-4)(n-5)}~
q_{m+l+k}-\frac{8(n-2)}{(n-2)(n-3)(n-4)(n-5)}~Q_{k,l,m}^3 \; .  \nn
\end{eqnarray}
Where we have introduced the following definitions:
\be
Q_{k,l,m}^3\equiv\int dq_1 dq_2 dq_3 P(q_1,q_2,q_3) q_1^lq_2^k
q_3^m  \; \;  \mbox{and}  \; \; 
Q_{k,l,m,p}^4\equiv\int dq_1 dq_2 dq_3 P(q_a,q_b,q_c,q_d) q_a^lq_b^k
q_c^m q_d^p \; .
\label{eqdef}
\ee
Finally we get the following expression for the term (\ref{eqq2qqb}): 
\begin{eqnarray}
\label{eq15b}
\overline{\lan q^2 \ran\lan q\ran^2 }&=&
\;\frac{(n-4)(n-8)+8(n-1))(n-1)^2}{(n-2)(n-3)(n-4)(n-5)}  q_1^2 q_2\\
&+& \frac{2(n-8)(n-1)}{(n-2)(n-3)(n-4)(n-5)}\left( q_2^2+2 q_1
q_{3}\right) \nn\\ 
&+& \frac{16}{(n-2)(n-3)(n-4)(n-5)}
q_{4}-\frac{8(n-2)}{(n-2)(n-3)(n-4)(n-5)}~Q_{1,1,1}^3 \; .  \nn
\end{eqnarray}
\end{itemize}

For the other term , $\overline{\lan q\ran^4 }$, calculations require 
more effort since one has two deal with terms involving four overlaps, but
the procedure is similar. We only quote the final result: 

\begin{eqnarray}
\label{eq16b}
\overline{\lan q\ran^4 } & = & \frac{((n-1)^3(-672+n(208 +(n-24) n))}
{(n-2)(n-3)(n-4)(n-5)(n-6)(n-7)}~q_1^4  
+ \frac{12 (n-1)^2(112 +(n-20)n)} {(n-2)(n-3)(n-4)(n-5)(n-6)(n-7)}q_1^2
~q_2  \\
&-& \frac{32(n-12)(n-1)}{(n-2)(n-3)(n-4)(n-5)(n-6)(n-7)} q_1\left(-2
q_3 +(n-2) ~Q_{1,1,1}^3\right) \nn\\
&+& \frac{ 12 ((n-16)
(n-1)}{(n-2)(n-3)(n-4)(n-5)(n-6)(n-7)} ~q_2^2 
+ \frac{ 20}{(n-2)(n-3)(n-4)(n-5)(n-6)(n-7)} ~q_4 \nn \\
&-& \frac{4(n-2)}{(n-2)(n-3)(n-4)(n-5)(n-6)(n-7)}
\left(8 ~Q_{2,1,1}^3- (n-3) ~Q_{1,1,1,1}^4 \right)\; .  \nn
\end{eqnarray}
Further calculations lead to a rather intricated  expression for $G_c$ in
the limit $n\to 0$:  
\be
G_c=\frac{49 q_1^4 - 77 q_1^2 q_2 + q_2^2 + 60 q_1 q_3 - 18 q_4 
+ 18 Q_{2,1,1}^3 - 24 q_1 ~Q_{1,1,1}^3 - 9  ~Q_{1,1,1,1}^4}{133  q_1^4 -
329 q_1^2 q_2 - 38 q_2^2 + 240 q_1 q_3 - 51 q_4 - 54 ~Q_{2,1,1}^3 + 72 q_1 
Q_{1,1,1}^3 + 27 Q_{1,1,1,1}^4 } \; . 
\label{G_c}
\ee
>From which the following results for $A_c$ and $B_c$ are obtained:
\be
A_c=\frac{-49 q_1^4 + 77 q_1^2 q_2 - q_2^2 - 60 q_1 q_3 + 18 q_4 - 18
~Q_{2,1,1}^3 +  
        24 q_1 ~Q_{1,1,1}^3 + 9  ~Q_{1,1,1,1}^4}{70(q_2-q^2)^2 }\; , \; 
\label{A_c}
\ee
\be
B_c=3\frac{91 q_1^4 - 203 q_1^2 q_2 + 34 q_2^2 + 80 q_1 q_3 - 17 q_4 - 18
~Q_{2,1,1}^3 +  
        24 q_1 ~Q_{1,1,1}^3 + 9  ~Q_{1,1,1,1}^4}{140(q_2-q^2)^2 } \; . 
\label{B_c}
\ee
Thus, one has to go to particular structures of the replica matrix $Q_{ab}$
to try to obtain a simpler expression for $B$, $A$, $G_c$, $B_c$, and
$A_c$.  
\vskip 0.1in
{\bf The symmetric case}\\
\vskip 0.1in
In the symmetric case, $B$, the Binder cumulant takes the value $1$ in the
frozen phase and  $A$ vanishes since there are  no OPF.  In all the
connected quantities, as well as in $G$, both numerator and denominator
vanish, thus RE gives no information. Despite, we have to stress that, as
shown numerically in RS for the spherical SK model, $G$
still takes the value $1/3$ in the SG phase.\\ 
\vskip 0.1in
{\bf One-step replica symmetry breaking}\\
\vskip 0.1in
In this particular case the structure of the replica matrix is the
following: replicas are distributed into $m$ blocks, so that  the matrix
elements can have two possible values $q_1$ if the two replicas belong to
the same block  and $q_0$ if they belong to different blocks. Thus, we have
for the different terms which appear in the calculation of $G_c$, $A$,
$A_c$, $B$ and $B_c$: 

\be
\frac{1}{n(n-1)}\sum_{a,b} Q_{ab}^k=\frac{(m-1)
q_1^k+(n-m)q_0^k}{n-1} \; ,\\ 
\label{eq17b}\ee
\bea
\label{eq18b}
\frac{1}{n(n-1)(n-2)}&\sum_{a,b,c}&
Q_{ab}Q_{bc}Q_{ca}= \\
&& \frac{q_1(m-1)[q_1^2(m-2)+q_0^2(n-m)]+q_0(n-m)[2
q_1q_0(m-1)+q_0^2(n-2m)]}{n-1}~, \nn
\eea
\bea
\label{eq19b}
\frac{1}{n(n-1)(n-2)}&\sum_{a,b,c}&
Q_{ab}^2Q_{bc}Q_{ca}= \\ 
&&\frac{q_1^2(m-1)[q_1^2(m-2)+q_0^2(n-m)]+q_0^2(n-m)[2
q_1q_0(m-1)+q_0^2(n-2m)]}{n-1}\; , \nn
\eea
\bea
\label{eq20b}
\frac{1}{n}\sum_{a,b,c,d} Q_{ab} Q_{bc} Q_{cd}
Q_{da}&=&(m-1)((m-3)(m-2)q_1^4 - 2(3 m-5) (m - n) q_1^2 q_0^2 +4 (2 m^2 -
3 m n + n^2) q_1 q_0^3 )\nn \\
&& -(3 m^2 + m (2 - 3 n) + (n-1)^2) (m - n) q_0^4 \; .
\eea

Substituting all these previous expressions into relations (\ref{eq16b}),
(\ref{eq14b}), (\ref{eq15b}) yields, in the limit $n\to 0$ the expression
already given in section III for the connected quantities $G_c$
(\ref{onestepgc}), $A_c$ (\ref{onestepac}) and $B_c$ (\ref{onestepbc}) : 
\be
G_c=\frac{39 - 113 m + 98 m^2}{93 - 221 m + 266 m^2}\; .
\ee     
In the same way we obtain expressions for $A_c$ and $B_c$ which do not
depend on the actual values of the elements of the replica matrix: 
\be
A_c=\frac{39 - 113 m + 98 m^2}{140 m (1-m)}\; \mbox{and} \; B_c=\frac{3(31
- 167 m + 182 m^2)}{280 m (1-m)}\; . 
\ee     
For the other disconnected quantities, $A$ and $B$ we obtain an expression
which depends not only on $m$ but also on the elements of $Q_{ab}$, this
is,  $q_0$ and $q_1$. The expressions are the following: 
\be
A=-\frac{( m-1)\;m\;(q_1^2 - q_0^2)^2}{3 (( m-1)\; q_1^2 - m\;
q_0^2)^2}\; , 
\label{onestepa}
\ee     
\be
B=\frac{( m-1) \;q_1^4 - m\; q_0^4 +  3 (( m-1)\; q_1^2 - m \;q_0^2)^2}{2
(( m-1)\; q_1^2 - m\; q_0^2)^2 
}\; . 
\label{onestepb}
\ee     
Comparing $A,B$ to the other connected parameters we find a further
dependence on the parameters $q_0,q_1$. Computations for further
steps of RSB are performed in the same way. However for a two-step patter
for $Q_{ab}$, we do not obtain a simple expression from which it is possible
to extract general conclusions.

\hspace{-2cm}

\end{document}